\documentclass{emulateapj}
\shorttitle{Accretion of chondrules in the solar nebula}
\shortauthors{C.W. Ormel, J.N. Cuzzi and A.G.G.M. Tielens}
\usepackage[varg]{txfonts}
\newcommand{\se}[1]{\mbox{\S\ref{sec:#1}}}
\newcommand{\Se}[1]{\mbox{\S\ref{sec:#1}}}
\newcommand{\eq}[1]{\mbox{equation\ (\ref{eq:#1})}}
\newcommand{\eqp}[1]{\mbox{eq.\ [\ref{eq:#1}]}}
\newcommand{\Eq}[1]{\mbox{Equation\ (\ref{eq:#1})}}
\newcommand{\fg}[1]{\mbox{Fig.\ \ref{fig:#1}}}
\newcommand{\Fg}[1]{\mbox{Figure\ \ref{fig:#1}}}
\newcommand{\Tb}[1]{\mbox{Table\ \ref{tab:#1}}}
\newcommand{\ie}{i.e.,}
\newcommand{\cq}{c.q.,}
\newcommand{\eg}{e.g.,}
\begin{abstract}
We present a mechanism for chondrules to stick together by means of compaction of a porous dust rim they sweep up as they move through the dusty nebula gas. It is shown that dust aggregates formed out of micron-sized grains stick to chondrules, forming a porous dust rim.  When chondrules collide, this dust can be compacted by means of rolling motions within the porous dust layer. This mechanism dissipates the collisional energy, compacting the rim and allowing chondrules to stick. The structure of the obtained chondrule-dust agglomerates (referred to as compounds) then consists of three phases: chondrules, porous dust, and dust that has been compacted by collisions. Subsequently, these compounds accrete their own dust and collide with other compounds. The evolution of the compound size distribution and the relative importance of the phases is calculated by a Monte Carlo code. Growth ends, and a simulation is terminated when all the dust in the compounds has been compacted. Numerous runs are performed, reflecting the uncertainty in the physical conditions at the chondrule formation time. It is found that compounds can grow by 1-2 orders of magnitudes in radius, upto dm-sizes when turbulence levels are low. However, relative velocities associated with radial drift form a barrier for further growth. Earlier findings that the dust sweep-up by chondrules is proportional to their sizes are confirmed. We contrast two scenarios regarding how this dust evolved further towards the densely packed rims seen in chondrites.
\end{abstract}
\keywords{solar system: formation --- planetary systems: formation and protoplanetary disk}
\begin{document}
\title{Co-accretion of Chondrules and dust in the Solar Nebula}
\author{C.W. Ormel}
\affil{Kapteyn Astronomical Institute, University of Groningen\\
       P.O. Box 800, 9700 AV Groningen,\\ 
       The Netherlands}
\email{ormel@astro.rug.nl}
\and
\author{J.N. Cuzzi, A.G.G.M. Tielens}
\affil{NASA Ames Research Center \\ 
       Mail Stop 245-3,\\
       Moffett Field, CA 94035, USA}
\email{jcuzzi@mail.arc.nasa.gov, atielens@mail.arc.nasa.gov}
\section{Introduction}
\label{sec:intro}
Protoplanetary nebulae have been studied in increasing detail from visual to microwave wavelengths \citep{2007prpl.conf..573M,2007prpl.conf..523W}, and hundreds of extrasolar planetary systems have been discovered, but the `primary accretion' stage of the planetary formation process -- that which leads from interstellar grains to planetesimals large enough to decouple from the nebula gas (asteroid- and comet-nucleus size objects) -- remains obscure. In this particle size range, coupled particle-gas dynamics dominates the evolution, as reviewed recently by \citet{2005ASPC..341..732C}, \citet{2006mess.book..353C}, and \citet{2007prpl.conf..783D}. The main processes which have been hypothesized for primary accretion include \textit{i)} incremental growth by sticking of small grains to each other and to larger particles, \textit{ii)} various kinds of instabilities occurring in a particle-rich midplane region, and \textit{iii)} formation of planetesimals from dense zones of particles that form in turbulence due to vorticity or pressure effects.

A critical but unknown nebula property in this stage is whether turbulence is present, and if so, what its intensity is \citep{2000prpl.conf..589S,2005ASPC..341..145G}. If the nebula is nonturbulent, particles of all sizes can settle into a dense layer near the midplane where incremental growth is fairly robust for expected, but still poorly known and therefore somewhat ad-hoc, sticking properties \citep{1993prpl.conf.1031W,1993Icar..106..102C,1997Icar..127..290W,2000SSRv...92..295W,2004come.book...97W}. This is because the dense particle layer drives the entrained gas to corotate, and relative velocities between equal-size particles would largely vanish. The high local mass density ensures that growth is rapid \citep{2000SSRv...92..295W} -- perhaps too rapid \citep{2005ASPC..341..732C}. 
Various instabilities in such a layer, mostly gravitational, have been studied for decades \citep{1973ApJ...183.1051G,1998Icar..133..298S,2002ApJ...580..494Y,2005ApJ...620..459Y}, but these are precluded if the nebula is even weakly turbulent \citep{2006mess.book..353C}. 

Astronomical and planetary observations seem to be most naturally reconciled with turbulent nebulae \citep{2005A&A...434..971D,2005ASPC..341..732C,2006Brownlee_etal,2006Zolensky_etal,2007LPI....38.1386C}. A number of studies indicate that turbulence excites meter-size particles to relative velocities at which they probably disrupt each other \citep{1988mess.book..348W,2000SSRv...92..279B,2004Icar..167..431S,2007arXiv0711.2148L}, posing a barrier to further growth. However, some recent studies suggest that turbulence itself can concentrate particles of different sizes, in different ways, and trigger rapid planetesimal formation \citep{2001ApJ...546..496C,2007LPI....38.1439C,2006MNRAS.372L...9R,2007Natur.448.1022J}. Thus, in spite of the ongoing uncertainty in just how turbulence may be maintained \citep{2000prpl.conf..589S,2006ApJ...653..503M}, it is sensible to consider its effect in model studies. This collisional disruption limit, combined with the rapid inward drift of m-size particles by which they are `lost' from the local region, led to the concept of a `m-size barrier' or bottleneck to growth; once large particles exceed this barrier, relative velocities become lower, allowing them to grow further and to drift less rapidly out of the accreted region. Our studies were initially motivated by a desire to see if growth into loose fractal clusters and subsequent packing could allow the m-size barrier to be crossed.

In the conceptually simplest models, growth occurs by simple sticking of particles \citep{1997Icar..127..290W,2000SSRv...92..295W,2004come.book...97W,2004ASPC..309..369B}. While it goes against our earthbound intuition that macroscopic particles can stick to each other, some microgravity and earthbased experiments show that, while bouncing transpires at intermediate velocities, sticking prevails for both low (less than a meter/second) and high (for $13-25\ \mathrm{m\ s^{-1}}$) relative velocities \citep{2000Icar..143..138B,2004PhRvL..93k5503B,2005Icar..178..253W,2005GeoRL..3211202M,2007prpl.conf..783D}. Other experiments indicate that certain solids (water and methanol ice, organic material) are `stickier' than others (silicates) \citep{1996Icar..123..422B}. However, no significant amount of these especially sticky materials has been found in primitive meteorites. Still, entire chondrites (and by inference entire parent bodies) are composed of small silicate objects that seem to have been gently assembled and compacted, at least initially \citep{1992GeCoA..56.2873M,1996ASPC,2006mess.book..353C}; how did this happen? The meteorite record (discussed in more detail below) shows that many mm-sized solid objects are encased in rims of micron and submicron-sized mineral grains. One obvious possibility is that these rims form by nebula accretion of grains onto the underlying core particles \citep{1984GeCoA..48.2581N,1992GeCoA..56.2873M,1997LPI....28.1071P,2002M&PS...37..229H,2003GeCoA..67.1711Z}. In \se{model} we show that this dust accretion may be expected to occur fractally, leading to porous structures. Collisions easily crush this structure, in the process dissipating kinetic energy and allowing colliding particles to stick \citep{BlumWurmARAA2008}.

Recent reviews of the relevant properties of chondrites are provided by \citet{1998BrearleyJones}, \citet{2005ASPC..341...15S} and \citet{2006mess.book...19W}. Chondrites are dominated by mm-sized silicate chondrules, which were melted in the nebula \citep{1998BrearleyJones,2000prpl.conf..927J,2005ASPC..341..251J,2006mess.book.....L}, but are found in meteorites to be embedded in a fine-grained matrix. 
Formation of chondrules (and chondrites) occurred over a period of several Myr \citep{2005ASPC..341...15S,2005ASPC..341..558K,2006mess.book..233R,2006LPI....37.1884K}. Chondrites can be divided into three broad classes -- ordinary, carbonaceous and enstatite -- with each class being further subdivided into more than a dozen groups reflecting chemical, mineralogical and isotopic differences. For example, CM carbonaceous chondrites contain abundant matrix that has been affected by aqueous alteration, while Ordinary Chondrites contain very little matrix and have generally incurred only limited aqueous alteration. Violent collisional processes occurred after primary accretion which affected the contents and appearance of most meteorites, and to best understand the primary accretion process one must look back through this stage where possible to the rare, unbrecciated subset of rocks and rock fragments called `primary texture' \citep{1992GeCoA..56.2873M,1993GeCoA..57.1521B}.

The dust in chondrites is found to have two physically defined components: rims and inter-chondrule matrix \citep{1981GeCoA..45...33H,1988mess.book..718S,1996ASPC,1998BrearleyJones}. Fine-grained rims are clearly associated with individual chondrules and other macroscopic particles in microscopic images, and usually even stay attached to the chondrules when they are disaggregated from the host rock \citep{1997LPI....28.1071P}. Some studies report that the composition of these fine-grained rims is uniform across a wide range of underlying mineral types, including more refractory (higher-temperature) oxides which formed much earlier \citep{1993GeCoA..57.1521B,2002M&PS...37..229H} and some find dramatic variations between the composition of rims on adjacent chondrules \citep{1983chto.conf..262T,1984GeCoA..48.1741S}. Generally, the chondrules cooled completely before accreting these fine dust grains \citep{1993GeCoA..57.1521B}. Interchondrule matrix, more generally dispersed between all the macroscopic components of the rock, is also made of fine-grained material. The grain sizes in fine-grained rims are noticeably smaller than the ubiquitously enveloping matrix, even though the compositions of the rims and matrix are very similar or identical \citep{1977E&PSL..35...25A,1993GeCoA..57.1521B,1996ASPC,1993GeCoA..57.3123Z}. It has been reported that the rim porosity is also smaller than that of the surrounding matrix \citep{1977E&PSL..35...25A,2006GeCoA..70.1271T}. The relative abundance of rim and matrix material to chondrule material varies from one meteorite class to another \citep{1988mess.book..718S}; however, the rim mass (or thickness) is often found to be proportional to the mass (or radius) of the underlying chondrule \citep{1992GeCoA..56.2873M,1997LPI....28.1071P}. 

Several different model efforts have attempted to explain some of these properties in the context of nebula sweep-up or accretion of the fine-grained rims by chondrules and their like. \citet{1998Icar..134..180M} hypothesized that if a particle had a speed relative to the gas which was proportional to its radius, and if the chondrules in a region sweep up all the local dust in a one-stage event (no ongoing replenishment of dust), the observed rim-core correlation would be obtained. \citet{2004Icar..168..484C} showed that chondrule-size particles in turbulence plausibly exhibit just the appropriate (near-linear) dependence of relative velocity on size, even though most particles obey a square-root dependence on radius. \citet{2004Icar..168..484C} relied on collisional outcomes proposed by \citet{1997ApJ...480..647D} for porous aggregates of fine grains, and suggested that for particles much larger than chondrules, the velocity relative to the gas increases to a point where they enter an erosional regime.

On the other hand, \citet{1993Metic..28..669S} and \citet{2006GeCoA..70.1271T} question whether fine-grained rims are nebula accretion products at all. \citet{2006GeCoA..70.1271T} point out in particular that the fine-grained rims in CM chondrites, such as discussed by \citet{1992GeCoA..56.2873M}, have a porosity of 10-20\%, much lower than the high-porosity structures formed by, \eg\ \citet{2000Icar..143..138B} or \citet{2004PhRvL..93k5503B}. Less is known quantitatively about the porosity of fine-grained rims in other chondrite types, although \citet{1977E&PSL..35...25A} states that rim porosities are less than 6-15\% in ordinary chondrites. The alternate that \citet{2006GeCoA..70.1271T} and \citet{1993Metic..28..669S} prefer, while they differ in the details, is that the fine-grained rims seen in CM chondrites, in particular, are created on the parent body from a generic enveloping matrix, by some combination of compaction and pervasive aqueous alteration. This suggestion might make it harder to explain why the grain size is smaller than in the nearby enveloping matrix. Nevertheless, the discussion shows that the porosity of fine-grain rims is an important diagnostic of their origin.

In this paper, we develop a detailed collision model to study the rimming and accretion processes of chondrules simultaneously and, in a statistical study, quantify the growth that can be obtained under a wide range of (uncertain) nebular conditions. Our model treats multiple components: solid `chondrules,' submicron grains and their very porous nebula aggregates, porous accretion rims formed by direct accretion of monomers and aggregates onto chondrules, compact rims formed by collisional compression of pre-existing porous rims, and compound objects formed by sticking of rimmed objects, which themselves might become rimmed in dust. Our collisional outcomes use physical guidelines set by laboratory and theoretical models \citep{1997ApJ...480..647D,2000Icar..143..138B,2004PhRvL..93k5503B,2007arXiv0711.2148L}. We use quantitatively correct closed form relative velocity expressions for particles in turbulence of varying intensity \citep{2007A&A...466..413O}, which capture the increase in relative velocity as particles grow by accretion of other particles. We use a Monte Carlo approach to calculate the probability of different outcomes, over a wide range of nebula parameters (level of turbulence, gas and solid density). We assess \textit{i)} the extent to which fine-grained rims can dissipate collisional energy and allow growth by sticking to proceed, and \textit{ii)} the extent to which these dissipative collisions compact initially porous dust rims into lower porosity states. We leave for future study the physics of disruptive collisions and the details of vertically varying particle density and turbulent intensity, such as might occur if the global turbulent intensity is vanishingly small \citep{1993Icar..106..102C,1998Icar..133..298S,1999JGR...10430805D,1997Icar..127..290W}.

This paper is organized as follows. In \se{model} the collision model is discussed. Here we outline the three distinct components with which we model the compound objects that result out of accretion of dust on chondrule surfaces and collisions with other chondrules. We also introduce the different sources of relative velocities particles can obtain in the nebula, and calculate the timescales involved in the various accretion processes. We end this discussion with a brief summary of the envisioned collisional scenario. \Se{mccoag} briefly reviews the Monte Carlo code with which the coagulation is solved. \Se{results} presents the results of our work. First, a few individual models are addressed in detail, before we present the results of a parameter study in which many uncertain (mostly nebula-related) parameters are varied. In \se{discuss} we discuss the effects of a particle dominated environment caused by settling of compounds on the growth of compounds. We also discuss several observational implications, emphasizing in particular the relation between the dust in our model to the fine-grained rims seen around chondrules in meteorites. We summarize our results in \se{summ}.

\section{Model}
\label{sec:model}
\begin{deluxetable}{lp{70mm}}
  \tablecaption{\label{symbollist} List of frequently used symbols }
  \tablehead{\colhead{Symbol} & \colhead{Description}}
  \startdata
    $\Delta v$ or $\Delta v_{12}$   & relative velocity \\
    $\Omega$                        & local Keplerian orbital frequency  \\
    $\alpha$                        & turbulent strength parameter (\se{physcon})   \\
    $\gamma$                        & surface energy density   \\
    $\delta$                        & fractal growth parameter (\se{dustaggr}) \\
    $\epsilon$                      & size ratio ($\epsilon \le 1$)                \\
    $\eta$                          & nebula pressure parameter (\eqp{etadef}) \\
    $\lambda$                       & mean free path (gas) (\se{physcon})   \\
    $\nu_\mathrm{m}, \nu_\mathrm{T}$ & molecular/turbulent kinematic viscosity (\se{physcon}) \\
    $\phi, \phi_\mathrm{PCA}, \phi_\mathrm{pd}$& filling-factor (PCA/porous dust) (\Tb{phases})   \\
    $\sigma$                        & collisional cross section  \\
    $\tau_\mathrm{f}$               & friction time (eqs.\ [\ref{eq:tauEp}] and [\ref{eq:tauSt}])   \\
    $\rho_\mathrm{X}$               & gas density over MSN (\se{physcon}) \\
    $\rho_\mathrm{d}, \rho_\mathrm{g}$ & spatial dust/gas density (\se{physcon})   \\
    $\rho_\mathrm{c}^\mathrm{(s)}, \rho_\mathrm{d}^\mathrm{(s)}$  & specific material chondrule/dust density (\Tb{phases}) \\
    $\rho_1, \rho_2$                & internal particle density \\
    $C_{ij}$                        & collision rate between particles $i,j$ (\se{mccoag}) \\
    $E$                             & collisional energy (\se{aquis}) \\
    $E_\mathrm{roll}, E_\mathrm{br}$ & monomer rolling/breakup energy (\eqp{stickroll}) \\
    $H_\mathrm{g}$                  & gas scaleheight (\se{turbmot})\\
    $R$                             & heliocentric radius      \\
    $\mathrm{Re}$                   & Reynolds number (\se{physcon}) \\
    ${\cal R}_\mathrm{gc}$          & spatial gas-chondrule mass-ratio (\se{physcon})   \\
    ${\cal R}_\mathrm{cd}$          & spatial chondrule-dust mass ratio (\se{physcon})   \\
    $\mathrm{St}$                   & particle Stokes number (\se{systmot})   \\
    $P$                             & pressure              \\
    $T$                             & temperature              \\
    $V$                             & geometrical (total) compound volume (\eqp{Vgeodef}) \\
    $v_\mathrm{g}, v_\mathrm{pg}$   & gas and particle-gas turbulent velocity (\se{turbmot}) \\
    $a$                             & geometrical (total) radius (\se{systmot})   \\
    $a_\mu$                         & reduced radius, $a_1a_2/(a_1+a_2)$  \\
    $a_\mathrm{dust}$               & dust aggregate radius (\se{dustaggr})  \\
    $a_\mathrm{ch}$                 & chondrule radius      \\
    $a_\mathrm{0}$                  & monomer radius  (\se{physcon}) \\
    $c_\mathrm{g}$                  & sound speed (gas) (\se{physcon}) \\
    $f_\mathrm{comp}$               & required dust fraction at collision for sticking (\eqp{fcoll})\\
    $f_\mathrm{d}$                  & compound dust fraction by mass (\Tb{phases}) \\
    $f_\mathrm{geo}$                & geometry factor  (\se{dissip}) \\
    $f_\mathrm{p}$                  & compound porous dust fraction with respect to total dust mass (\Tb{phases}) \\
    $m,m_\mathrm{d},m_\mathrm{ch}$  & (dust/chondrule) mass \\
    mw-.. or $\langle .. \rangle_m$ & mass-weighted averages (see \eqp{massavdef}) \\
    $t_\mathrm{dd}, t_\mathrm{dc}, t_\mathrm{cc}$ & dust-dust/dust-chondrule and chondrule-chondrule collision times (\se{evolstruc}) \\
    $t_\mathrm{L}/t_\mathrm{s}$     & large/small eddy turn-over time (\se{turbmot}) \\
    $v_\mathrm{K}$                  & orbital (Kepler) velocity             \\
    $v_\mathrm{r}$                  & particle radial drift velocity  (\eqp{vrad})
  \enddata
  \tablecomments{\strut List of frequently used symbols that re-occur over various sections in the paper.}
\end{deluxetable}

\subsection{Outline\label{sec:outline}}
The central theme of this paper is to model the process of dust accretion onto chondrule surfaces and explore whether compaction of this dust during inter-chondrule collisions acts as a sticking agent, with which significant growth can be achieved. For dust aggregates this compaction mechanism is well known \citep{1997ApJ...480..647D,2000Icar..143..138B,2007ApJ...661..320W}: by restructuring of the constituent grains the excess collisional energy is dissipated. For dust-rimmed chondrules we argue the situation is analogous, except that part of the aggregate's interior is now replaced by a chondrule. The prerequisite for such a scenario is the presence of a reservoir of dust that is accreted fractally by the chondrules, preserving its fluffy structure. Here, we follow the \citet{1998Icar..134..180M} `closed box' scenario in which a fixed amount of dust is injected instantaneously to the chondrule population. The compound objects (or, simply, compounds) thus obtained are modeled to consist of three phases: chondrules, compact (\ie\ restructured) dust, and porous (\ie\ fractally accreted) dust. The restructuring mechanism also holds for collisions between compounds, again at the expense of the porous phase. In this way a coagulation process is initiated by which chondrules are accreted into large compounds. This coagulation is only stopped when the compounds run out of porous dust, such that the collisional energy can no longer be dissipated; at this stage, all the free-floating dust has been accreted and no more porous dust remains. 

The environment in which these processes take place is a key ingredient that enters the coagulation model. A violent, low dust density environment leads to high velocity collisions which quickly pack down the porous rim, limiting its capability to allow further sticking, while even higher velocities will lead to break-up of compounds. On the other hand, if (relative) velocities are modest and remain so during the phase in which compounds accrete other compounds and porous dust, this could lead to significant growth. In this work we use various sources for relative velocities: thermal, turbulent and systematic. The relative velocity is further determined by the internal structure (density) of the compounds, which affects their coupling to the gas. The internal structure of compounds is reflected in the definition of the `geometric size': the size that corresponds to the effective aerodynamic cross section. The evolution of the internal structure during the accretion process is therefore of key importance: \ie\ do collisions follow `hit-and-stick' behaviour in which growth proceeds fractally, or do collisions keep the filling factor constant.

\subsection{The turbulent nebula\label{sec:nebula}}

\subsubsection{Disk physical structure and model components\label{sec:physcon}}
\begin{deluxetable}{llll}
  \tablecaption{\label{tab:physpar} Gas and dust parameters}
  \tablehead{  \colhead{(1)} & \colhead{(2)} & \colhead{(3)} &\colhead{(4)} }
  \startdata
    gas density$^\star$       & $\rho_\mathrm{g}$                                 & $28/ 2.4/ 0.16$             & $10^{-11}\ \mathrm{g\ cm^{-3}}$ \\
    sound speed               & $c_\mathrm{g}$                                    & $10./ 7.6/ 5.6$             & $10^4\ \mathrm{cm\ s^{-1}}$  \\
    mean free path (gas)      & $\lambda$                                         & $6.9/ 82/ 1230$             & $ \mathrm{cm}$          \\
    temperature               & $T$                                               & $280/ 162/ 89$              & $ \mathrm{K}$           \\
    pressure parameter        & $\eta$                                            & $1.6/2.7/4.9$               & $ \times10^{-3}$           \\
    large eddy turn-over time & $t_\mathrm{L} = \Omega^{-1}$                      & $0.16/ 0.82/ 5.0$           & $ \mathrm{yr}$ \\
    turbulent strength parameter$^\star$   & $\alpha$                            & $10^{-4}$                                      \\
    inner eddy turn-over time$^\star$ & $t_\mathrm{s}$                                    & $1.3/12/131$                & $10^3\ \mathrm{s}$\\
    \hline
    gas-chondrule ratio$^\star$& ${\cal R}_\mathrm{gc}$                           & $100$                                   \\
    gas-dust ratio$^\star$    & ${\cal R}_\mathrm{cd}$                            & $100$                                   \\
    chondrule mean size       & $a_\mathrm{ch}$                                   & $300$                         & $\mu\mathrm{m}$\\
    monomer dust size         & $a_0$                                             & $1$                           & $\mu\mathrm{m}$                             \\
    fractal growth parameter    & $\delta$                                        & $0.95$                                         \\
    surface energy density dust$^\star$ & $\gamma$                                        & $19$                          & $\mathrm{erg\ cm^{-2}}$ 
  \enddata
  \tablecomments{Parameters characterizing the gas (upper rows) and dust/chondrules (bottom rows). Gas parameters correspond to a minimum mass solar nebula (MSN) model of total gas mass of $2.5\times 10^{-2}\ \mathrm{M}_\odot$ inside 10 AU with power law exponents of, respectively, $-1.0$ and $-0.5$, for the surface density and temperature structure as function of radius \citep{2002ApJ...581.1344T}. Columns denote: (1) parameter description; (2) symbol, ${\cal R}_\mathrm{gc} = \rho_\mathrm{g}/\rho_\mathrm{c}$ and ${\cal R}_\mathrm{cd} = \rho_\mathrm{c}/\rho_\mathrm{d}$; (3) corresponding value with multiple values denoting conditions at 1, 3 and 10 AU, respectively (3 AU is the default); (4) unit. Parameters indicated by a $\star$ are variables (default model values are given).}
\end{deluxetable}
The start of the model is defined as the point at which populations of chondrules and dust interact. Before this point, the two populations were either spatially isolated or one did not exist. This paper is not concerned with the history of the two populations -- specifically, we avoid the nagging chondrule formation question -- but merely define the zero time of the model ($t=0$) as the point where the two populations mix. Of course, the history of the two populations determines to a large extent the conditions that prevail at the start of the model, \eg\ the size of the dust particles. We take $1\ \mu\mathrm{m}$ as the monomer size (\ie\ the smallest constituent size of a dust grain), which roughly corresponds to the sizes of the fine grains observed in chondrites \citep{1977E&PSL..35...25A}. These grains may have formed by condensation onto seed grains \citep{2004ASPC..309..213C} and could have aggregated into larger (fluffy) dust particles before they interact with the chondrules. For instance, if dust condenses out in a region of (recently formed) chondrules \citep{2004LPI....35.2140W,2005ApJ...623..571S}, the size distribution will be dominated by monomers. However, if chondrules encounter a dust cloud only after a certain time since its formation ($\sim 10^{2..3}\ \mathrm{yr}$), larger aggregates are expected to have formed through monomer collisions. In this work, though, the size of the dust aggregates ($a_\mathrm{dust}$) is simply treated as a free model parameter (\se{evolstruc}). 

Both solid components find themselves in the gas-dominated protoplanetary disk. We use a minimum mass solar nebula (MSN) model \citep{1981PThPS..70...35H} to determine the gas parameters (Table\ \ref{tab:physpar}, after \citealt{2002ApJ...581.1344T}). At $R = \mathrm{3\ AU}$ this corresponds to a spatial density of $\rho_\mathrm{g} = 2.4\times10^{-11}\ \mathrm{g\ cm^{-3}}$, a thermal speed of $c_\mathrm{g} = 7.6\times10^4\ \mathrm{cm\ s^{-1}}$ and a mean free path of $\lambda = 82\ \mathrm{cm}$. Studies of chondrule formation indicate, however, that gas densities may be much higher (\eg\ \citealt{2002M&PS...37..183D,2006Natur.441..483C}) and we will therefore treat $\rho_\mathrm{g}$ as a free parameter, denoting with $\rho_X$ the density enhancement over MSN. We further assume the gas surface density ($\Sigma$) and temperature ($T$) profiles are power-laws of heliocentric radius ($R$) and fix the exponents at $-1.0$ and $-0.5$, respectively. This choice is consistent with a steady disk in which $\dot{\Sigma}=0$ and an accretion flow, $\dot{M}$, independent of radius.\footnote{Provided the $\alpha$-turbulence model (\eqp{turbnu}) is assumed and boundary conditions are neglected (see, \eg\ \citealt{1981ARA&A..19..137P}).} The gas-to-chondrule and the chondrule-to-dust density ratios, 
\begin{equation}
  {\cal R}_\mathrm{gc} = \rho_\mathrm{g}/\rho_\mathrm{c}; \textrm{ and } {\cal R}_\mathrm{cd} = \rho_\mathrm{c}/\rho_\mathrm{d},
  \label{eq:Rdef}
\end{equation}
are free parameters; for example, ${\cal R}_\mathrm{cd}=1$ means that chondrules and dust are present in equal proportion (as observed in some meteorite groups, \citealt{2005ASPC..341..701H}) and ${\cal R}_\mathrm{gc}=100$ is the standard gas to solids ratio. In our model chondrules follow a log-normal size distribution, for which we take parameters of $300\ \mu\mathrm{m}$ (the mean) and a (log-normal) width parameter of $0.5$ (see, \eg\ \citealt{1984Metic..19..135R,2002M&PS...37.1361N}). This makes the mean, mass-weighted size of the chondrule population, $\langle a \rangle_m = 720\ \mu\mathrm{m}$. Note that the initial distribution is not necessarily equivalent to the distribution that ends up in meteorites, or the distribution extracted from meteorites by the thin section method \citep{1996M&PS...31..243E}. 
An overview of all parameters characterizing the gas and solids is given in \Tb{physpar}.

We assume that the gas in the disk is in a turbulent state of motion. 
After \citet{1973A&A....24..337S}, the turbulent viscosity is parameterized as
\begin{equation}
  \nu_\mathrm{T} = \alpha c_\mathrm{g} H_\mathrm{g} = \alpha c_\mathrm{g}^2 / \Omega,
  \label{eq:turbnu}
\end{equation}
where $H_\mathrm{g}$ is the scaleheight of the gas disk, $\Omega$ the local (Keplerian) rotation velocity, and $\alpha$ a scale parameter that determines the strength of the turbulence \citep{1973A&A....24..337S}. Values for $\alpha$ are very uncertain.  If the magneto-rotational instability is active it may be up to $10^{-3}$ \citep{1991ApJ...376..214B,1991ApJ...376..223H}; however in regions of low ionization it can be much lower \citep{1996ApJ...457..355G,2000ApJ...543..486S}. The extent of the turbulence is determined by the Reynolds number, $\mathrm{Re}$, defined as $\mathrm{Re} = \nu_\mathrm{T}/\nu_\mathrm{m}$, with $\nu_\mathrm{m}$ the molecular viscosity, $\nu_\mathrm{m}=c_\mathrm{g} \lambda/2$ \citep{1993Icar..106..102C}. The turbulent spectrum consists of eddies characterized by a scale ($\ell$), velocity ($v$), and turn-over time ($t$), between an outer (or integral) scale $L$ and an inner (or Kolmogorov) scale $\ell_\mathrm{s}$. Following previous works, $t_\mathrm{L}$, the largest eddy turn-over time, is taken equal to the inverse orbital frequency, $t_\mathrm{L} = 1/\Omega$, and $v_\mathrm{L}=\alpha^{1/2}c_\mathrm{g}$ (\eg\ \citealt{1995Icar..114..237D,2001ApJ...546..496C,2004ApJ...614..960S}). The eddy properties at the turbulence inner scale then follow from the Reynolds number:
\begin{equation}
  t_\mathrm{s} = \mathrm{Re}^{-1/2} t_\mathrm{L};\qquad \ell_\mathrm{s} = \mathrm{Re}^{-4/3} L;\qquad v_\mathrm{s} = \mathrm{Re}^{-1/4} v_\mathrm{L}.
  \label{eq:turbss}
\end{equation}

\subsubsection{Thermal motions\label{sec:thermal}}
When gas molecules collide with a larger (dust) particle momentum is transferred, changing the motion of the dust particle. These kicks occur stochastically, resulting in a velocity behaviour known as Brownian motion. The ensuing velocity difference between two particles of mass $m_1$ and $m_2$ is highest for low masses and high temperatures,
\begin{equation}
  \Delta v_\mathrm{BM} = \sqrt{\frac{8k_\mathrm{B}T(m_1+m_2)}{\pi m_1 m_2}},
  \label{eq:Brownian}
\end{equation}
where $k_\mathrm{B}$ is Boltzmann's constant. For micron-sized particles Brownian velocities are a few $\mathrm{mm/s}$; but since $\Delta v_\mathrm{BM}$ decreases with the $-3/2$ power of the size of the smallest particle, it quickly becomes negligible for larger particles.

\subsubsection{Systematic motions\label{sec:systmot}}
The key parameter that determines the coupling of solids to the gas is the friction time, $\tau_\mathrm{f}$. In the Epstein regime the size of the particle, $a$, is small with respect to the mean-free-path of gas molecules, $\lambda$, and the friction time is given by
\begin{equation}
  \tau_\mathrm{f} = \tau_\mathrm{f}^\mathrm{Ep} = \frac{3}{4c_\mathrm{g}\rho_\mathrm{g}}\frac{m}{\pi a^2}. \qquad (a\le \case{9}{4}\lambda)
  \label{eq:tauEp}
\end{equation}

For solid $1\ \mu\mathrm{m}$-grains the friction time is $\tau_\mathrm{f} \sim 10^2/\rho_X\ \mathrm{s}$ for the default nebula parameters of 3 AU (\Tb{physpar}), while for an $a\sim 300\ \mu\mathrm{m}$ chondrule it takes $\sim 10/\rho_X$ hours before the traces of its initial motion are `erased.' Note that \eq{tauEp} defines $a$ as the geometrical radius of the particle, \ie\ the radius corresponding to the angularly-averaged projected surface area of the particle. If the particle is a fluffy aggregate its friction time is therefore much less than a compact-equivalent with the same mass. If significant growth takes place, particles will no longer obey the Epstein drag law; 
friction times are then enhanced with respect to $\tau_\mathrm{f}^\mathrm{Ep}$, \ie\
\begin{equation}
  \tau_\mathrm{f}^\mathrm{St} = \frac{4 a}{9\lambda} \frac{1}{\mathrm{Re_p} C} \tau_\mathrm{f}^\mathrm{Ep}.\qquad (a \ge \case{9}{4}\lambda)
  \label{eq:tauSt}
\end{equation}
Here, $\mathrm{Re_p} = 2av_\mathrm{pg}/\nu_\mathrm{m}$ is the \textit{particle} Reynolds number, which determines the constant $C$,\footnote{Compared to, \eg\ \citet{1977MNRAS.180...57W} the definition of $C$ has been scaled down by a factor of 24. That factor is already present in \eq{tauSt}.} and $v_\mathrm{pg}$ the particle-gas velocity. 
Within the physical conditions of the simulations in \se{results} the particle Reynolds number stays below $\mathrm{Re_p}=1$, for which $C=\mathrm{Re_p^{-1}}$ \citep{1977MNRAS.180...57W}. Friction times are then independent of $v_\mathrm{pg}$.

One of the well-known problems in the planet-formation field is the strong inward radial drift particles of a specific size experience, \eg\ meter-sized particles at $\sim$AU radii or cm-sized at $\sim 100$ AU radii  (see \citet{2007A&A...469.1169B} for a recent review). This inward radial drift is caused by the existence of gas pressure gradients, resulting in a gas velocity that is somewhat less than Keplerian by a difference of magnitude $\eta v_\mathrm{K}$ \citep{1977MNRAS.180...57W,1986Icar...67..375N} with $\eta$ the dimensionless pressure parameter, defined as
\begin{equation}
  \eta \equiv - \frac{1}{2R\Omega^2} \frac{1}{\rho_\mathrm{g}} \frac{\partial P}{\partial R} \approx c_\mathrm{g}^2/v_\mathrm{K}^2.
  \label{eq:etadef}
\end{equation}
However, particles do not experience this pressure term and instead attempt to move at Keplerian velocities, faster than the gas. The ensuing drag force removes angular momentum from the particle resulting in a radial velocity of \citep{1977MNRAS.180...57W}\footnote{In \eq{vrad} we have not accounted for collective effects when the particle density is comparable to or higher than the gas density. \Eq{vrad} then changes \citep{1986Icar...67..375N}. Angular momentum exchange between the dust and gas dominated layers \citep{2004ApJ...601.1109Y} is another process to be accounted for, but its significance is relatively modest \citep{2007A&A...469.1169B}.}
\begin{equation}
  v_\mathrm{r} = - \frac{2\mathrm{St}}{1+\mathrm{St}^2} \eta v_\mathrm{K},
  \label{eq:vrad}
\end{equation}
where we have defined $\mathrm{St} = \tau_\mathrm{f}\Omega$. This systematic radial drift velocity peaks at $\mathrm{St}=1$. Chondrule-sized particles, however, are generally sufficiently well coupled to the gas ($\mathrm{St} \ll 1$) so that radial motions are low in most physical conditions; turbulent-induced motions then dominate (unless $\alpha$ is really low). However, when particles grow in size, systematic motions may take over from turbulent velocities (see \fg{motions}).

\begin{figure}
  \plotone{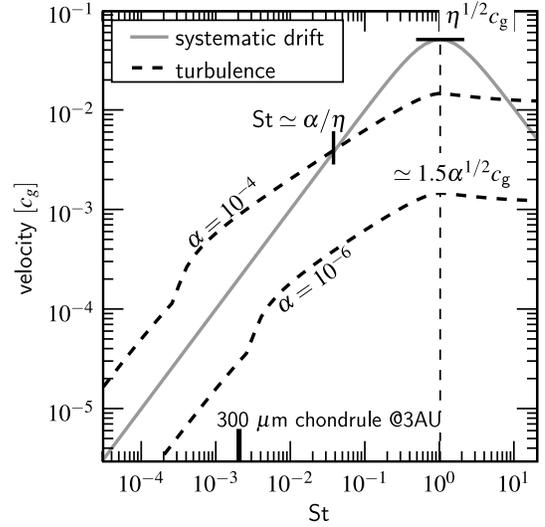}
  \caption{\label{fig:motions}Comparison of systematic and turbulent velocities as function of particle Stokes number. Plotted are radial velocities (\eqp{vrad}, solid grey curve) and turbulent velocities with $\varepsilon=0$ (\eqp{relvs}, dashed curves). Note the change in slope between the linear and the square-root turbulent velocity regime which happens at $\mathrm{St} \simeq \mathrm{Re}^{-1/2}$. Turbulent velocities dominate over systematic velocities for $\mathrm{St} \lesssim \alpha/\eta$ (provided this is $\gtrsim \mathrm{Re}^{-1/2}$). The Stokes number corresponding to a $300\ \mu\mathrm{m}$-sized chondrule at the default nebular conditions of 3 AU (see Table\ \ref{tab:physpar}) is also indicated. All velocities peak at $\mathrm{St}=1$ (light-dashed vertical curve).}
\end{figure}
\subsubsection{Turbulent motions\label{sec:turbmot}}
For a Kolmogorov spectrum, turbulence leads to mean (large scale) velocity fluctuations of $v_\mathrm{g} = (3/2)^{1/2}v_\mathrm{L} = (3/2)^{1/2} \alpha^{1/2} c_\mathrm{g}$ \citep{2003Icar..164..127C}. Due to their inertia, solids do not instantaneously follow these fluctuations but require a time $\tau_\mathrm{f}$ before their motions align. This leads to a net relative motion, $v_\mathrm{pg}$, between the gas and the solid particle of \citep{2003Icar..164..127C}
\begin{equation}
  v_\mathrm{pg} = v_\mathrm{g} \sqrt{\frac{\mathrm{St}^2(\mathrm{Re}^{1/2}-1)}{(\mathrm{St}+1)(\mathrm{St}\mathrm{Re}^{1/2}+1)}}, 
  \label{eq:Vpgturb}
\end{equation}
where the Stokes number, $\mathrm{St}$, is the ratio between the friction time and the large eddy turn-over time, \ie\ $\mathrm{St}=\tau_\mathrm{f}/t_\mathrm{L} = \tau_\mathrm{f}\Omega$.\footnote{$\mathrm{St}=\tau_\mathrm{f}/t_\mathrm{L}$ is the formal definition for the Stokes number. In the $\alpha$-turbulence model $t_\mathrm{L} = \Omega^{-1}$ and the definitions for $\mathrm{St}$ in equations (\ref{eq:vrad}) and (\ref{eq:Vpgturb}) coincide.} The limiting expressions of \eq{Vpgturb}, $v_\mathrm{pg} = \mathrm{St}v_\mathrm{g}$ for $\mathrm{St}\ll1$ and $v_\mathrm{pg} = v_\mathrm{g}$ for $\mathrm{St} \gg 1$, respectively, correspond to the cases of particles that are well coupled (small particles) and poorly coupled (larger particles) to the gas. 

The calculation of particle-particle relative velocities does not follow directly from the $v_\mathrm{pg}$'s since particle velocities can become very incoherent in turbulence (\eg\ $\Delta v_{12} \neq | v_\mathrm{1g} - v_\mathrm{2g}|$). Consider, for example, two small particles entrained in the same eddy. If their motions are aligned, no relative velocity is present; it is only within a time $\tau_\mathrm{f}$ after being caught in the eddy that these particles have the chance to develop relative motions, provided their friction times differ (\ie\ $\tau_\mathrm{1} \neq \tau_\mathrm{2}$). The problem of finding suitable (\ie\ closed-form) expressions for $\Delta v_{12}$ is important since these are key to any model of dust coagulation (\eg\ \citealt{1997Icar..127..290W,2001ApJ...551..461S,2005A&A...434..971D,2006ApJ...640.1099N}) including this work. Following earlier works of \citet{1980A&A....85..316V}, \citet{1984Icar...60..553W}, \citet{1991A&A...242..286M} and \citet{2003Icar..164..127C}, \citet{2007A&A...466..413O} have presented closed-form analytical expressions for $\Delta v_{12}$ (with a margin of error of $\sim 10\%$) in terms of the Stokes number of the particles:
\begin{equation}
\left( \frac{\Delta v_{12}}{v_\mathrm{g}} \right)^2 = 
  \cases{
  \mathrm{Re}^{1/2} \left( \mathrm{St}_1 - \mathrm{St}_2 \right)^2                                                                                                                       & for $\tau_1 < t_\mathrm{s}$ \cr
  \left[ 2y_\mathrm{a}^* - (1+\varepsilon) + \frac{2}{1+\varepsilon} \left( \frac{1}{1+y_\mathrm{a}^*} + \frac{\varepsilon^3}{y_\mathrm{a}^*+\varepsilon} \right) \right] \mathrm{St}_1  & for $5t_\mathrm{s} \simeq \tau_1 \lesssim t_\mathrm{L}$ \cr
  \left( \frac{1}{1+\mathrm{St}_1} + \frac{1}{1+\mathrm{St}_2} \right)                                                                                                                   & for $\tau_1 \ge t_\mathrm{L}$
  }
\label{eq:relvs}
\end{equation}
In these expressions $\tau_1$ (or $\mathrm{St}_1$) always corresponds to the particle of the largest friction time and $\varepsilon = \tau_2/\tau_1 \leq 1$. Near the $\tau_1 = t_\mathrm{s}$ turning point the expression for $\Delta v_{12}$ is somewhat more complex (see \citealt{2007A&A...466..413O}). $y_\mathrm{a}^*$ is a numerical constant of value $y_\mathrm{a}^* \simeq 1.60$ if $\tau_1 \ll t_\mathrm{L}$; however, when $\tau_1 \simeq t_\mathrm{L}$ it becomes a function of $\tau_1$ and drops to unity at $\tau_1=t_\mathrm{L}$. In that case we approximate $y_\mathrm{a}^*$ by an interpolation function. In \fg{motions} $\Delta v_{12}$ is plotted for two values of $\alpha$ in the limit of $\mathrm{St}=\mathrm{St}_1 \gg \mathrm{St}_2$ (dashed curves). The three regimes of \Eq{relvs} are clearly distinguishable: the linear regime for $\tau_\mathrm{f} \lesssim t_\mathrm{s}$ (or $\mathrm{St} \lesssim \mathrm{Re}^{-1/2}$; small particles); the square-root regime, $t_\mathrm{s} \lesssim \tau_\mathrm{f} \lesssim t_\mathrm{L}$; and the high Stokes regime, $\mathrm{St} \ge 1$, where particles decouple from the gas. Small particles like chondrules have Stokes number $\mathrm{St} \ll 1$; whether they fall into the linear or square-root velocity regime depends on their sizes in relation to the gas parameters (\eg\ $\alpha, \rho_\mathrm{g}$) of the disk. 

\Fg{motions} also shows the systematic drift velocity (\eqp{vrad}, solid curve). If one assumes that $\mathrm{St}_2 \ll \mathrm{St}_1$ the radial drift curve also gives the relative velocity a particle with $\mathrm{St}=\mathrm{St}_1$ has with a much smaller particle. Actually, for $\mathrm{St}_1 \le 1$ turbulent and systematic drift relative velocities (eqs.\ [\ref{eq:vrad}] and [\ref{eq:relvs}]) depend only weakly on $\mathrm{St}_2$ (the lower Stokes number); the curves in \fg{motions} can therefore be interpreted as the typical relative velocity a particle of a given Stokes number has with particles of similar or lower Stokes numbers. \Fg{motions} shows that for very small particles ($\tau_\mathrm{f} \ll t_\mathrm{s}$ or $\mathrm{St}\ll \mathrm{Re}^{-1/2}$) turbulent velocities only dominate when $\alpha \gtrsim 10^{-5}$. Then, when turbulent velocities flatten out in the square-root regime, the radial drift motion may catch up with the point of intersection lying at $\mathrm{St} \simeq \alpha/\eta$, provided $\alpha$ is not either too low or too high. For any model with $\alpha \lesssim \eta$ radial drift motions will eventually dominate: a regime of high relative velocities ($\sim\mathrm{10\ m\ s^{-1}}$) is therefore unavoidable.

\subsection{Collisions between dust-rimmed chondrules\label{sec:colmod}}
\begin{deluxetable}{llll}
  \tablecaption{\label{tab:phases}Material properties of compounds phases}
  \tablehead{  \colhead{Phase} & \colhead{Specific density} & \colhead{Mass fraction\tablenotemark{a}} &\colhead{Filling factor} }
  \startdata
  chondrule       & $\rho_\mathrm{c}^\mathrm{(s)} = \mathrm{3\ g\ cm^{-3}}$ & $(1-f_\mathrm{d})$                & $\phi_\mathrm{ch} =1$       \\
  compact dust    & $\rho_\mathrm{d}^\mathrm{(s)} = \mathrm{3\ g\ cm^{-3}}$ & $f_\mathrm{d} (1-f_\mathrm{p})$   & $\phi_\mathrm{cd}=0.33$  \\
  porous dust     & $\rho_\mathrm{d}^\mathrm{(s)} = \mathrm{3\ g\ cm^{-3}}$ & $f_\mathrm{d} f_\mathrm{p}$       & $\phi_\mathrm{pd} \lesssim 0.15$
  \enddata
 \tablenotetext{a}{with respect to entire compound (sum equals 1); $f_\mathrm{d} =$ mass fraction in dust; $f_\mathrm{p} = $ porous mass fraction of the dust.}
\end{deluxetable}

Theoretical studies and laboratory experiments have shown that the outcome of grain-grain or aggregate-aggregate collisions depends on its ratio of the kinetic energy to a critical energy \citep{1997ApJ...480..647D,2000Icar..143..138B}. Specifically, porous dust is accreted when chondrules sweep up dust grains, or aggregates of dust grains, at collisional energies ($E$) that stay below the energy threshold for restructuring, $5E_\mathrm{roll,}$ where $E_\mathrm{roll}$ is the energy required to roll one contact area over the surface of the grain.  This can lead to a very open structure of filling factors ($\phi_\mathrm{pd}$) that are lower than the filling factors obtained in particle-cluster aggregation, $\phi_\mathrm{PCA} \simeq 0.15$ (see below, \se{collcomp}). As outlined in \se{outline}, the absence of restructuring during dust sweep-up is crucial since the resulting porous structure can then be compacted in more energetic collisions --  the collision of compounds -- promoting further growth. In compound collisions, $E>E_\mathrm{roll}$ and the dust within the compound will restructure, dissipating a unit of $\sim E_\mathrm{roll}$ for each dust grain that is involved in the rolling motion.
The porous dust that is involved in restructuring compacts to a higher filling factor, $\phi_\mathrm{cd}$ (\se{collcomp}).

Compounds can then be represented as a three phase structure: chondrules, compact dust and porous dust. Two numbers, the dust fraction $f_\mathrm{d}$ and the porous dust fraction $f_\mathrm{p}$, quantify the relative importance of each phase within a compound (see \Tb{phases}). The internal structure of each phase is further characterized by its filling factor, $\phi$. A schematic picture of the structure is given in \fg{phases}. 

\begin{figure}
  \plotone{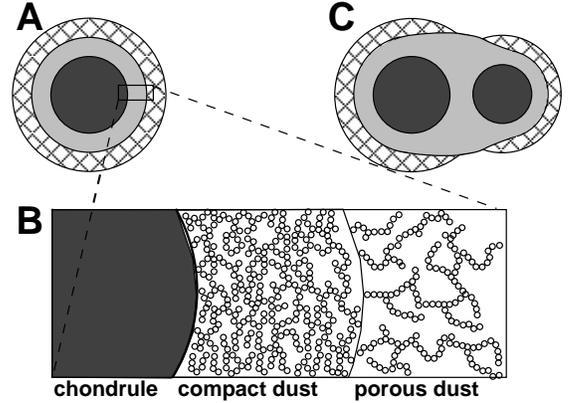}
  \caption{\label{fig:phases}(\textit{A}) Schematic representation of the three-phase model of dust-rimmed chondrules: chondrule (dark), compact dust (grey) and porous dust (pattern). The inset (\textit{B}) shows the substructure of the dust that consists of micron-sized monomers. (\textit{C}) If chondrules collide, the collision energy is dissipated by transferring dust from the porous to the compact dust phase. Figure not to scale.}
\end{figure}
\subsubsection{\label{sec:collcomp}Collisional compaction}
The accreted dust mantles surrounding chondrules can have a very porous and fractal structure. Typically, grains in these rims will be bonded to two other grains in large string-like structures. When two rimmed chondrules collide, contact will be established between two (or a limited number of pairs of) grains and these grains will bear the full brunt of the collision. Once the force on these grains exceeds the critical rolling force, they start to roll (restructuring). The rolling of these grains may enable contact formation between more pairs of grains, thereby promoting compaction and at the same time reducing the force per contact. Compaction will stop when the force on newly made contacts drops below the rolling force. Compaction may also stop because the resulting structure is too rigid to allow for further rolling, \ie\ the rolling grain made contact with too many grains. Since forces are propagated through such compacted structures, this means that none of the grains involved experiences a force exceeding the rolling force.  The compression of the contact area in a collision between two monomers will give rise to an elastic repulsion force slowing down and eventually reversing the collision. The absolute value of the repulsive force will be set by the kinetic energy of the collision; in the Hertzian limit the sum of the forces on the individual contacts scales with the remaining kinetic energy to the 3/5th power. For colliding aggregates, rolling of the contacting monomers  provides an additional energy dissipation channel. 
But once the resulting structure is too rigid to enable further rolling, compression of the individual contact areas will provide the repulsion required to slow down and possibly even reverse the collision if the collision is energetic enough.  The collision partners then bounce.

\begin{figure}
  \plotone{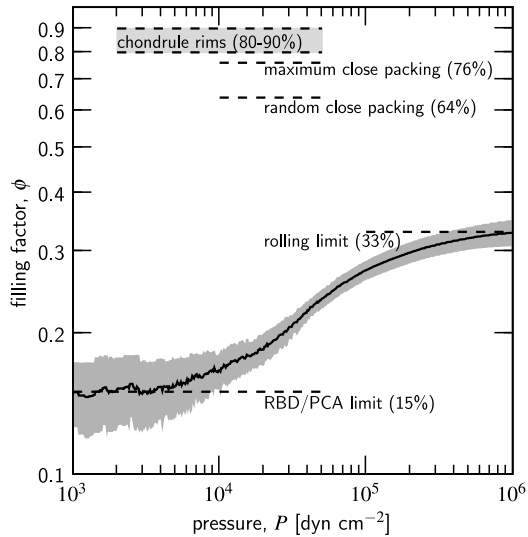}
  \caption{\label{fig:comprcurve}(solid curve) Compression of PCA aggregates, obtained by static compression of `dust cakes' created by random ballistic deposition of $a_0=0.75\ \mu\mathrm{m}$ SiO$_2$ spheres \citep{2004PhRvL..93k5503B,2006ApJ...652.1768B}. The uncertainty in the measurements is denoted by the grey area. Values for $\phi$ for several amounts of packing configurations are shown. (Data on the compression curve experiments kindly provided by J\"urgen Blum.)}
\end{figure}
\citet{2004PhRvL..93k5503B} have designed experiments to measure the compaction of dust cakes under uniaxial compression. These dust cakes were grown through a Particle-Cluster-Aggregation (PCA) method by deposition of individual $1.5\ \mu\mathrm{m}$ diameter monomers at low velocities where rolling is not a factor and growth occurs through a hit-and-stick process. The volume filling factor of the resulting aggregates was measured to be $0.15$, in good agreement with numerical simulations of this process \citep{1993A&A...280..617O,1997wats}. At this volume filling factor, the typical coordination number, \ie\ the number of neighbours with which the monomer is in contact, is calculated to be 2 \citep{1967ridgway}.  
These dust cakes were exposed to a unidirectional pressure in a static experiment. \Fg{comprcurve} shows the resulting volume filling factor as a function of the applied unidirectional pressure \citep{2004PhRvL..93k5503B}. The results show that compaction is initiated at an applied pressure of $\sim10^4\ \mathrm{dyn\ cm^{-2}}$.  If we assume that the number of monomers per unit area being pressed on is given by,
\begin{equation}
  N/A = \left( \pi a_0^2 \phi^{2/3} \right)^{-1} \sim 2\times10^8\ \mathrm{cm^{-2}}
  \label{eq:NoverA}
\end{equation}
with $a_0$ the radius of the monomer, the force on an individual monomer becomes $\sim 5\times10^{-5}\ \mathrm{dyn}$. This is very close to the rolling force of $7\times10^{-5}\ \mathrm{dyn}$ \citet{2004PhRvL..93k5503B} extrapolated from the measurements by \citet{1999PhRvL..83.3328H}. As the dust cake compacts and the average coordination number increases, the structure becomes more rigid and resistant to further compression (under these conditions, see below). Finally, at a pressure of $\sim 10^6\ \mathrm{dyn\ cm^{-2}}$ the structure is dense enough for rolling motions to be inhibited. This corresponds to an average coordination number of 3.9 and a filling factor of 0.33 (\fg{comprcurve}).

The conditions in the protoplanetary disk under which dust rims are formed by grain accretion and under which they evolve through collisions with other rimmed grains differ from those in these laboratory studies. First, the initial structure of the dust rims accreted on the chondrule surface may differ from PCA. Although low velocity collisions are expected, the monomers may collide preferentially among each other before colliding with a chondrule. In that case the resulting structure is referred to as Cluster-Cluster Aggregation (CCA), a process that leads to much lower filling factors than PCA. Whether PCA or CCA is preferred depends on the relative abundance of dust and chondrules and their relative velocities. So, the filling factor of the porous dust component may start lower than the experimental one in \fg{comprcurve}. However, since this process is directly tied to the rolling force experienced by the monomers that make contact, we expect that this difference in initial structure will have no influence on the critical pressure required for the onset of compaction. We expect, likewise, that uniaxial compression of dust rims grown by CCA will stall at 0.33 filling factor since this is again a property of the resulting structure; \eg\ at these kinds of volume filling factors, monomers in the dust rims will have been organized in `stabilizing' structures. 

However, under nebular conditions continuous impacts will arise from random directions; it is therefore likely that collisional compaction under these conditions will result in higher volume filling factors than the unidirectional compression experiments would indicate, possibly even as high as 0.5 (the value of $\phi=0.64$ characterizes for Random Close Packing, RCP). We note that \citet{2004PhRvL..93k5503B} and \citet{2006ApJ...652.1768B} in their compression studies did approach RCP when applying an omnidirectional pressure of $\sim10^9\ \mathrm{dyn\ cm^{-2}}$. Note, however, that omnidirectional pressure is not achieved in collisions between two bodies in an open environment; \ie\ the dust has the chance to spread perpendicular to the direction of compression, and the obtained high volume filling factor may not be generally attainable. Indeed, for $\phi>0.33$ rolling motion become impeded and we do expect that in order to `crush' the dust (rims) to RCP values much higher pressures are required. Studies indicate pressures of $\sim10^{10}\ \mathrm{dyn\ cm^{-2}}$ in order to reach RCP \citep{2003JMechPhysSolids...51..667,2005JSME...48..376}. This second stage of compaction would correspond to a very different collision regime characterized by much higher energies. Adopting the Hertzian limit, we expect that this higher pressure, \cq\ force in the contact area corresponds to a impact energy which is higher by a factor $(10^9/10^6)^{5/3}$ or a collision velocity higher by a factor $10^{5/2}$ over the velocity/energy required to initiate compaction.  In this study, while acknowledging that higher filling factors are plausible, we have for simplicity assumed that $\phi=0.33$ is the limiting value.

\subsubsection{Acquisition of a porous dust layer\label{sec:aquis}}
\begin{deluxetable}{lll}
  \tablecaption{\label{tab:rolstickpar} Critical energies}
  \tablehead{ \colhead{Expression} & \colhead{Breaking}  & \colhead{Rolling} }
  \startdata
   Theoretical    \tablenotemark{a}           & $E_\mathrm{br} = A_\mathrm{br} \gamma^{5/3} a^{4/3}_\mu / {\cal E}^{2/3} $ & $E_\mathrm{roll} = A_\mathrm{roll} \xi_\mathrm{crit} a_\mu \gamma$ \\
   DT97 Prefactors\tablenotemark{a}           & $A_\mathrm{br} \simeq 9.6$                                                     & $A_\mathrm{roll} \simeq 59$ \\
   Measured energies\tablenotemark{b}         & $E_\mathrm{br} = 1.3\times10^{-8}\ \mathrm{erg}$                              & $E_\mathrm{roll}=1.7\times10^{-8}\ \mathrm{erg}$ \\
   Empirical prefactors\tablenotemark{c}      & $A_\mathrm{br} = 2.8\times10^{3}$                                             & $A_\mathrm{roll} = 1.8\times10^{3}$
  \enddata
  \tablenotetext{a}{Theoretically derived expressions for $E_\mathrm{br}, E_\mathrm{roll}$ from \citet{1993ApJ...407..806C} (for the breakup energy) and \citet{1997ApJ...480..647D} (for rolling) and corresponding pre-factors, $A_\mathrm{br}, A_\mathrm{roll}$ . We define $\xi_\mathrm{crit} = 10^{-8}\ \mathrm{cm}$.}
  \tablenotetext{b}{Values adopted from \citet{2000Icar..143..138B} for parameters of $a_0 = 2a_\mu = 9.5\times10^{-5}\ \mathrm{cm}$, $\gamma = \mathrm{19\ erg\ cm^{-2}}$ and ${\cal E} = 3.7\times10^{11}\ \mathrm{dyn\ cm^{-2}}$. The original measurements were performed by \citet{1999AdSpR..23.1197P} (for the breakup energy) and \citet{1999PhRvL..83.3328H} (for the rolling energy).}
  \tablenotetext{c}{Empirically derived prefactors from the theoretical expressions with the measured values for $E_\mathrm{br}$ and $E_\mathrm{roll}$.}
  \tablecomments{Comparison between predicted and measured critical energies for breakup and rolling.}
\end{deluxetable}
Two critical energies -- the breakup and rolling energy -- regulate the behaviour of the dust (porous accretion/compaction) upon collision: \citep{1993ApJ...407..806C,1997ApJ...480..647D,2000Icar..143..138B}
\begin{eqnarray}
  \label{eq:stickroll}
    E_\mathrm{br} = A_\mathrm{br} \frac{\gamma^{5/3} a_\mu^{4/3}}{ {\cal E}^{2/3}}; \\
    \label{eq:Estick}
    E_\mathrm{roll} = 6\pi^2 \xi_\mathrm{crit} a_\mu \gamma = A_\mathrm{roll} \xi_\mathrm{crit} a_\mu \gamma,
    \label{eq:Eroll}
\end{eqnarray}
where $a_\mu = a_1a_2/(a_1+a_2)$ is the reduced radius of the collision partners, $\gamma$ the surface energy density of the material and ${\cal E}$ Young's elastic modulus (assuming the same materials collide). $\xi_\mathrm{crit}$ in the $E_\mathrm{roll}$ expression is some critical distance used to initiate rolling which \citet{1997ApJ...480..647D} assumed to be on the order of the atomic size, $\xi_\mathrm{crit} = 10^{-8}\ \mathrm{cm}$. Using these definitions the constants $A_\mathrm{br}$ and $A_\mathrm{roll}$ are dimensionless. \citet{2000Icar..143..138B} have experimentally determined the breakup and rolling energies (see Table\ \ref{tab:rolstickpar}) and found these to be higher than the \citet{1997ApJ...480..647D} theoretical predictions. However, apart from a scale factor, the \citet{2000Icar..143..138B} experiments agreed well with the \citet{1997ApJ...480..647D} model; that is, collisions can be separated into the regimes of perfect sticking, restructuring and fragmentation. We therefore apply the mechanism put forward by \citet{1997ApJ...480..647D} but use pre-factors ($A_\mathrm{br}, A_\mathrm{roll}$) from the experimental results (last row of Table\ \ref{tab:rolstickpar}). Note that for micron-sized particles the rolling and breakup energy are similar.

When two particles meet, direct sticking occurs if the collision energy, $E$, is dissipated at the first point of contact; \ie\ $E \le E_\mathrm{stick}$, where $E_\mathrm{stick}$ is related to the breakup energy as $E_\mathrm{stick} = 0.22 E_\mathrm{br}$ \citep{1997ApJ...480..647D}. 
Writing $E=\case{1}{2}m_\mu (\Delta v)^2$ with $m_\mu = m_1m_2/(m_1+m_2)$ the reduced mass and $\Delta v$ the relative velocity, the criterion $E \le E_\mathrm{stick}$ translates into a threshold velocity of
\begin{eqnarray}
\nonumber
 v_\mathrm{st} = \sqrt{ \frac{2E_\mathrm{stick}}{m_\mu} } = \sqrt{0.45A_\mathrm{br}} \frac{ \gamma^{5/6} a_\mu^{4/6}}{  {\cal E}^{1/3} m_\mu^{1/2}}\\ \nonumber
 = 0.33 A_\mathrm{br}^{1/2} \left(\rho_\mathrm{d}^\mathrm{(s)}\right)^{-1/2} \gamma^{5/6} N_\mu^{-1/2} a_\mu^{2/3} a_0^{-3/2} {\cal E}^{-1/3} \\ \nonumber
 = 35\ \mathrm{cm\ s^{-1}}\ N_\mu^{-1/2} \left( \frac{a_\mu}{a_0} \right)^{2/3} \left( \frac{a_0}{\mathrm{\mu m}} \right)^{-5/6} \left( \frac{\rho_\mathrm{d}^\mathrm{(s)}}{\mathrm{3\ g\ cm^{-3}}} \right)^{-1/2}\\
  \times \left( \frac{\gamma}{\mathrm{19\ erg\ cm^{-2}}} \right)^{5/6} \left( \frac{\cal E}{\mathrm{3.7\times10^{11}\ dyn\ cm^{-2}}} \right)^{-1/3},
         \label{eq:stick}
\end{eqnarray}
where we have assumed that like materials meet (\ie\ same $\gamma, {\cal E}$) , and where the reduced mass has been parameterized as $m_\mu=N_\mu m_0$ with $m_0=4\pi\rho_\mathrm{d}^\mathrm{(s)}a_0^3/3$ the mass of the (smallest) grain. Thus, $N_\mu=1/2$ for equal-size particles and $N_\mu=1$ for very different size particles.
\Eq{stick} shows that micron-sized silicate particles have no problem to stick to each other at velocities of $\sim 10\ \mathrm{cm\ s^{-1}}$. This also holds for collisions between $\mu$m-sized grains and chondrules since it is the reduced size $a_\mu$ that enters the equation. However, at higher velocities the grains will bounce off.

In collisions between chondrules ($a_0 \sim 300\ \mu\mathrm{m}$) the sticking velocity falls below $\sim \mathrm{cm\ s^{-1}}$, lower than the velocities between chondrules for most values of $\alpha$ (see \fg{motions}). Also, for chondrules, the assumption of a smooth, spherical surface on which the physics behind \eq{Estick} is based breaks down. Although surface roughness increase the sticking capabilities for $\mu\mathrm{m}$-sized grains \citep{2000ApJ...533..454P}, the asperities in chondrules are probably too large to favour sticking. 
However, we now expect the previously accreted porous dust layer to act as the sticking mechanism through a dynamic restructuring and compaction of the constituent grains (\se{dissip}).

Another important collision is between a chondrule (or compound) and a dust grain or dust aggregate. Here, the criterion for sticking \textit{without restructuring} is $E\le 5E_\mathrm{roll}$ \citep{1997ApJ...480..647D}. This translates into a velocity of (using the same substitutions as above)
\begin{eqnarray}
  \nonumber
 v_\mathrm{st,aggr} = \sqrt{\frac{10E_\mathrm{roll}}{m_\mu}} = 1.6\times10^2\ \mathrm{cm\ s^{-1}}\ N_\mu^{-1/2} \left( \frac{a_\mu}{a_0} \right)^{1/2} \left( \frac{a_0}{1\ \mu\mathrm{m}} \right)^{-1} \\
 \times \left( \frac{\rho_\mathrm{d}^\mathrm{(s)}}{3\ \mathrm{g\ cm^{-3}}} \right)^{-1/2} \left( \frac{\gamma}{19\ \mathrm{erg\ g^{-1}}} \right)^{-1/2},
 \label{eq:vstick2}
\end{eqnarray}
in which now $N_\mu>1$ roughly corresponds to the number of grains in the aggregate. Thus, at moderately low velocities small aggregates will hit-and-stick, preserving their porous structure.  This `hit-and-stick' behaviour will also be referred to as `fractal accretion.' However, when aggregates become large or $\alpha$ is high (\eg\ $\alpha \ge10^{-3}$ and $\rho_X=1$; \fg{motions}) some compaction is likely to occur.  However, in this study we have ignored this effect (for reasons of computational efficiency) and simply assumed that all dust accretion occurs fractally. Although invalid in a violent collisional environment, the consequences of this assumption are marginal as the porous dust on the chondrule surface is quickly compacted by colliding chondrules in any case (\se{dissip}).

\subsubsection{\label{sec:dissip}Collisions between dust-chondrules compounds}
\citet{1991Icar...89..113H} studied collisions between cm-sized particles and found that sticking forces increased significantly when a frosty layer was present. While they attribute this enhanced sticking to interlocking of jagged surface structures, this effect probably reflects energy dissipation due to restructuring. In the case of chondrules, rimmed by a layer of fluffy dust, the situation is analogous: the fluffy structure allows the collisional energy to be dissipated away. Assuming that each monomer (of mass $m_0$, size $a_0$ and internal density $\rho_\mathrm{d}^\mathrm{(s)}$) in the porous dust layer is capable of absorbing an energy $E_\mathrm{roll}$, $E/E_\mathrm{roll}$ monomers are needed to dissipate the total collision energy. Expressed in terms of mass, a porous mass fraction of at least $f_\mathrm{comp}$ must be available, with $f_\mathrm{comp}$ the ratio of the required mass in porous dust to the total mass of the collision partners,
\begin{eqnarray}
  \nonumber
  f_\mathrm{comp} = \frac{m_0 E/E_\mathrm{roll}}{m_1+m_2} = \frac{4\pi}{3A_\mathrm{roll} \xi_\mathrm{crit}} \rho_\mathrm{d}^\mathrm{(s)} a_0^2 \gamma^{-1} \left( \frac{m_\mu^{2}}{m_1 m_2} \right) (\Delta v)^2 \\ \nonumber
= 3.7\times10^{-2} \left( \frac{m_\mu^{2}}{m_1 m_2} \right) \left( \frac{a_0}{\mathrm{\mu m}} \right)^{2} \left( \frac{\Delta v}{10\ \mathrm{cm\ s^{-1}}} \right)^2 \\
                  \times \left( \frac{\rho_\mathrm{d}^\mathrm{(s)}}{\mathrm{3\ g\ cm^{-3}}} \right) \left( \frac{\gamma}{\mathrm{19\ erg\ cm^{-2}}} \right)^{-1}.
  \label{eq:fcoll}
\end{eqnarray}
\begin{figure}
  \plotone{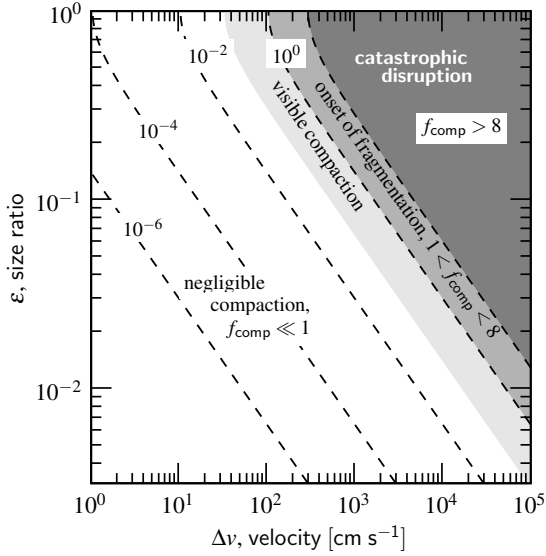}
  \caption{\label{fig:fcomp} Contours of $f_\mathrm{comp}$ (dashed curves) -- the fraction of the (combined) compound mass that must be involved in restructuring to dissipate away the collisional energy to stick the compounds (\eqp{fcoll}) -- as function of collision velocity ($x$-axis) and size ratio ($y$-axis). Equal internal densities are assumed, $\rho_1=\rho_2=3\ \mathrm{g\ cm^{-3}}$ and $a_0=1\ \mu\mathrm{m}$ The criterion for sticking is $f_\mathrm{comp} \le \langle f_\mathrm{d}f_\mathrm{p} \rangle_m$ (see text). For low velocities or size-ratio's compaction is insignificant. Collisions with $f_\mathrm{comp} \gtrsim \langle f_\mathrm{d}f_\mathrm{p} \rangle_m$ compact all their dust. When $f_\mathrm{comp} \gtrsim 8$ complete destruction occurs (see \se{impfrag}).}
\end{figure}
This equation reveals a few important results. First, $f_\mathrm{comp}$ decreases with smaller dust grains (smaller $a_0$); although more monomers are required to dissipate the same collision energy, the total mass of the monomers that restructures is less. Also, the dependence on velocity is rather steep; at very low velocities the amount of compacted material is negligibly low ($f_\mathrm{comp} \ll 1$), while visible compaction happens in a quite restricted velocity band (see \fg{fcomp}). Another important point is the mass-dependence in \eq{fcoll}. $f_\mathrm{comp}$ does not depend on the absolute masses of the particles involved, but, through the $m_\mu^{2}/m_1m_2$ factor, rather on the \textit{mass-ratio} of the collision partners. Thus, a collision between particles of very unequal size has a lower $f_\mathrm{comp}$ than equally-sized particles colliding at the same velocity and, therefore, a higher probability to stick (see \fg{fcomp}). This is of course due to the reduced mass that enters the collision energy. 

\subsubsection{\label{sec:recipe}Collisional recipe}
We will now quantify how the collisions affect the structural parameters of the compound, \ie\ the $f_\mathrm{d}$ and $f_\mathrm{p}$ quantities. \Eq{fcoll} gives the mass fraction of the collision partners that must be compacted, which, for sticking, must be less than the mass fraction available in porous dust ($f_\mathrm{d}f_\mathrm{p}$), averaged over the collision partners, \ie\ $\langle f_\mathrm{d}f_\mathrm{p}\rangle_m = (m_1f_{\mathrm{d},1}f_{\mathrm{p},1} + m_2f_{\mathrm{d},2}f_{\mathrm{p},2})/(m_1+m_2)$. If $f_\mathrm{comp} \le \langle f_\mathrm{d}f_\mathrm{p} \rangle_m$ enough porous dust is present to absorb the collisional energy and the two compounds stick. A fraction $f_\mathrm{comp}$ is then transferred from the porous to the compact phase. If $f_\mathrm{comp}>\langle f_\mathrm{p} f_\mathrm{d}\rangle_m$, however, restructuring cannot dissipate all the collisional energy. As mentioned before, we do not include other energy dissipation channels but simply consider all collisions in which $f_\mathrm{comp} > \langle f_\mathrm{d} f_\mathrm{p} \rangle_m$ to result in bouncing; $f_\mathrm{p}$ is then set to 0 for both particles. This means fragmentation of compounds or erosion of the porous rim are neglected (but see below, \se{impfrag}). 

A further restriction to the amount of dust that can be compacted is made when we account for the geometry of the collision. Then, only a fraction ($f_\mathrm{geo}$) of the compound (and of its dust) is involved in being compacted and dissipating energy. We estimate $f_\mathrm{geo}$ from the intersection between the particles' trajectories. This intersection actually is one between a cylinder and a sphere, but here we approximate it as a 2d intersection between two circles which meet at an impact parameter $b$. The area of the intersection, $A$, can be calculated by elementary geometry as (see \fg{project})
\begin{figure}[tbp]
  \plotone{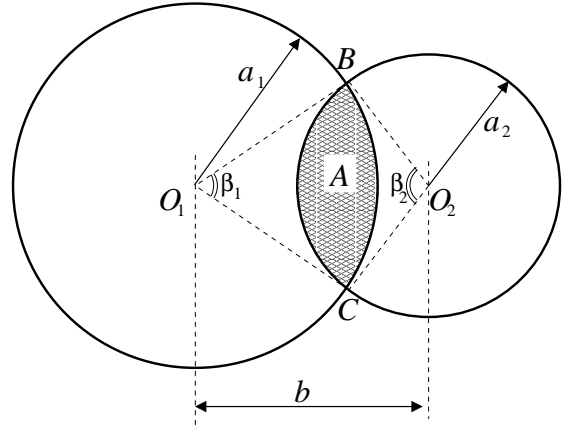}
  \caption{\label{fig:project}Projection of two compounds colliding at an impact parameter $b$. The ratio of the shaded region ($A$) relative to the cross section of each compound ($\pi a_i^2$) determines the fraction of the porous dust that can be used during the collision, \ie\ $f_{\mathrm{geo},i} = A/\pi a_i^2$. $A$ is obtained by subtracting the kite O$_1$BO$_2$C from the two circular sectors spanned up by $\beta_1$ and $\beta_2$ (\eqp{intersect}).}
\end{figure}
\begin{eqnarray}
  \nonumber
  A(a_1, a_2, b) = 2a_1^2 \arccos \left( \frac{b^2 + a_1^2 - a_2^2}{2ba_1} \right) + 2a_2^2 \arccos \left( \frac{b^2 + a_2^2 - a_1^2}{2b a_2} \right) \\
  - \sqrt{(-b+a_1+a_2)(b+a_1-a_2)(b-a_1+a_2)(b+a_1+a_2)},
\label{eq:intersect}
\end{eqnarray}
from which $f_{\mathrm{geo},i}$ for each particle is calculated as $f_{\mathrm{geo},i} = A/\pi a_i^2$. \Eq{intersect} is valid for impact parameters of $|a_1-a_2| < b < a_1+a_2$. For $b\le|a_1-a_2|$ the intersection equals the projected area of the smaller particle, while $A=0$ for $b\ge a_1+a_2$. The total mass-fraction of the particles that participates is $(m_1f_\mathrm{d1}f_\mathrm{p1}f_\mathrm{geo,1} + m_2f_\mathrm{d2}f_\mathrm{p2}f_\mathrm{geo,2})/(m_1+m_2)$ and this has to be greater than $f_\mathrm{comp}$ for sticking. Note, however, that inclusion of the $f_\mathrm{geo}$ factor might be too restrictive: since the sound speed inside aggregates ($\sim 30\ \mathrm{m\ s^{-1}}$; \citealt{2008Paszun}) is usually much higher than $\Delta v$, the energy will be quickly transferred along movable structures. For completeness we therefore consider both extremes: $f_\mathrm{geo}$ determined from \eq{intersect} and $f_\mathrm{geo}=1$.

When the collision results in sticking, the total dust and porous dust fractions are calculated as follows,
\begin{eqnarray}
    \label{eq:nwfd}
    f_\mathrm{d} = \frac{f_{d1} m_1 + f_{d2} m_2}{m_1+m_2}; \\
    f_\mathrm{p} = \frac{f_{d1} f_{p1} m_1 + f_{d2} f_{p2} m_2 - f_\mathrm{comp}(m_1+m_2)}{f_\mathrm{d}(m_1+m_2)}.
    \label{eq:nwfp}
\end{eqnarray}
For example, in \eq{nwfp} the three terms in the numerator denote, respectively, the mass in porous dust of particle 1, the porous dust mass of particle 2, and the porous mass transferred to the compact dust phase. In order to get the relative amount of porous dust this is divided by the new total dust mass (calculated in \eq{nwfd}) of the compound. If the collision results in a bounce, $f_\mathrm{d}$ stays the same for both particles and $f_\mathrm{p}$ is reduced by a factor $1-f_\mathrm{geo}$.

\subsubsection{\label{sec:impfrag}Role of fragmentation}
As the above formulas for $f_\mathrm{d}$ and $f_\mathrm{geo}$ suggest, fragmentation is not included in the collision model. The large number of particles produced by a fragmenting collision is especially problematic in the context of a Monte Carlo program, where the number of particles is limited (see \se{mccoag}). However, the results of \citet{1997ApJ...480..647D} provide insight into the stage at which fragmentation becomes important. They found that the onset of fragmentation starts at $E \simeq 0.3 N_\mathrm{c} E_\mathrm{br}$ with catastrophic disruption at energies of $\simeq 10 N_\mathrm{c} E_\mathrm{br}$, where $N_\mathrm{c}$ is the total number of contacts (roughly equal to the number of monomers, $N$) in an aggregate. Recalling from \se{aquis} that $E_\mathrm{br} \sim E_\mathrm{roll}$ the catastrophic fragmentation limit corresponds to $f_\mathrm{comp} \sim 8$ in terms of \eq{fcoll}. The corresponding curve in \fg{fcomp} then shows that fragmentation becomes important at velocities above a few $10^2\ \mathrm{cm\ s^{-1}}$ (for equal sized particles; as the mass disparity increases the fragmentation velocity increases). The m/s transition for the onset of fragmentation is in agreement with previous studies \citep{1993Icar..106..151B,2000Icar..143..138B} but compact structures at high filling factor may require more effort to fragment than their fluffy counterparts (D. Paszun, priv. comm.). From \fg{motions} the critical velocity can be translated into a Stokes number. We will \textit{a posteriori} check in which models fragmentation is expected to play a dominant role.

\subsection{Evolution of the internal structure\label{sec:evolstruc}}
The aerodynamic properties of the compounds, which determine their coupling to the gas, alter with accretion of dust and mutual collisions between compounds. These properties are quantified in the friction time, $\tau_\mathrm{f}$ (see \eqp{tauEp}), of the particles  -- essentially a measure of its mass-to-surface area ratio, $m/\pi a^2$. It is an important parameter since lower friction times mean lower relative velocity between the particles (\se{nebula}), and low relative velocities between compounds, in turn, imply potential to grow large. 

Through the adopted three phase model of compounds (\se{colmod}) the total geometrical volume, $V_\mathrm{geo}$, defined by $a$, can be reduced to its three components: \textit{i)} chondrule, \textit{ii)} compact dust, and \textit{iii)} porous dust, \ie\
\begin{equation}
  \frac{4\pi}{3} a^3 \equiv V = V_\mathrm{ch} + V_\mathrm{cd} + V_\mathrm{pd}.
  \label{eq:Vgeodef}
\end{equation}
In \se{colmod} the filling factors of the chondrule and compact dust phases have been fixed at 1 and 0.33, respectively, so that $V_i$ for these phases linearly corresponds to the mass inside these phases. However, for the porous dust phase, this does not have to be the case: the dust fluff-balls that are accreted can be of different size and porosity. Besides, if the compound accretion process itself proceeds fractally, the porous phase becomes a mixture of porous dust and voids created by the hit-and-stick packing of the compounds. This could lead to a much reduced filling factor of the porous phase (see \se{ccstick}).

With these issues in mind we discuss the three accretion modes that are at work and outline the implications for the internal structure of dust and compounds. In a largely chronological order these are: \textit{i)} dust-dust, \textit{ii)} chondrule-dust, and  \textit{iii)} compound-compound accretion.

\subsubsection{Dust-dust aggregation\label{sec:dustaggr}}
If the dust distribution initially consists of monomers of size $a_0$, the number density of dust particles is likely to be much larger than that of chondrules. Therefore, monomers probably form aggregates \textit{before} they are themselves accreted by chondrules or compounds. Provided the collisional energies involved stay below $5E_\mathrm{roll}$ (see \se{aquis}), the dust aggregates will hit-and-stick, leading to fractal growth \citep{1988ApJ...329L..39M,2000Blum_etal}. In addition, if this collision is between similar-size aggregates, the outcome is referred to as cluster-cluster aggregation (CCA). In that case fractal aggregates form with a surface area that grows faster with mass than in the compact case, i.e, $\pi a_\mathrm{dust}^2 \propto m^\delta$, or, $V \propto m^{3\delta/2}$ with $\delta = 0.95$ \citep{1993A&A...280..617O,2006Icar..182..274P} as compared to $V\propto m^{1/3}$ (or $\delta=2/3$) for compact growth. Using the relation $V=V_0 N^{3\delta/2}$, where $V_0$ denotes the volume of a single monomer and $N$ the number of monomers the aggregate contains, the filling factor evolves as 
\begin{equation}
  \label{eq:phidust}
  \phi_\mathrm{dust} = \frac{NV_0}{V} = \left( \frac{V}{V_0} \right)^{2/3\delta -1} = \left( \frac{a_\mathrm{dust}}{a_0} \right)^{2/\delta -3} = N^{1 - 3\delta/2}.
\end{equation}

We consider two mechanisms through which dust aggregates can form: \textit{i)} Brownian motion and \textit{ii)} differential velocities due to turbulence.
For simplicity, equal particle sizes are assumed at all times. The timescales involved are determined by the particle number density ($n_\mathrm{d}$), size ($a_\mathrm{dust}$) and relative velocities ($\Delta v$) between the particles, \ie
\begin{equation}
  \label{eq:tdust} 
  t_\mathrm{dd} = (n_\mathrm{d} \sigma \Delta v)^{-1} = \frac{1}{3}\frac{\rho_\mathrm{d}^\mathrm{(s)}}{\rho_\mathrm{d}}\frac{a_\mathrm{dust} \phi}{\Delta v},
\end{equation}
in which we have used $n_\mathrm{d} = \rho_\mathrm{d}/m$, $\sigma = \pi(a_1+a_2)^2 = 4\pi a_\mathrm{dust}^2$ for the collisional cross section, and $m=4\pi a_\mathrm{dust}^3 \phi \rho_\mathrm{d}^\mathrm{(s)}/3$ for the mass of a dust aggregate, with $\rho_\mathrm{d}^\mathrm{(s)}$ the specific material density of the dust. The relative velocities in the case of Brownian motion and turbulence read, respectively,
\begin{eqnarray}
    \Delta v^\mathrm{BM} = \sqrt{\frac{8k_\mathrm{B}T}{\pi m_\mu}} = \frac{2}{\pi} \sqrt{\frac{3k_\mathrm{B}T}{\rho_\mathrm{d}^\mathrm{(s)}}} a_\mathrm{dust}^{-3/2} \phi^{-1/2}; \\
    \Delta v^\mathrm{T} = \frac{v_\mathrm{s}}{t_\mathrm{s}} \frac{\rho_\mathrm{d}^\mathrm{(s)}}{c_\mathrm{g}\rho_\mathrm{g}} a_\mathrm{dust}\phi,
    \label{eq:vTurb}
\end{eqnarray}
If $\tau_\mathrm{f}<t_\mathrm{s}$, turbulent relative velocities are in fact determined by the spread within their friction times (see \se{turbmot}) and a numerical factor of, \eg\ 0.1 in front of \eq{vTurb} can be inserted if the particle distribution is monodisperse. This expression is further reduced by writing $v_\mathrm{s}/t_\mathrm{s} = \mathrm{Re}^{1/4} \alpha^{1/2} c_\mathrm{g} \Omega$. For Brownian motion, the timescale $t_\mathrm{dd}$ then becomes
\begin{equation}
  \label{eq:tcBrown}
  t_\mathrm{dd}^\mathrm{BM}  = \frac{\pi(\rho_\mathrm{d}^{(s)})^{3/2} {\cal R}_\mathrm{gd} }{6\rho_\mathrm{g}\sqrt{3k_BT}} \phi^{3/2} a^{5/2} \approx 1.4\times10^2\ \mathrm{yr}\ \rho_X^{-1} \phi^{3/2} \left( \frac{a_\mathrm{dust}}{1\ \mu\mathrm{m}} \right)^{5/2},
\end{equation}
and for turbulence, 
\begin{equation}
  \label{eq:tcturb}
  t_\mathrm{dd}^\mathrm{T} \approx 10\times \frac{1}{3} {\cal R}_\mathrm{gd} \mathrm{Re}^{-1/4} \alpha^{-1/2} \Omega^{-1} \approx 6\times10^2\  \mathrm{yr}\ \rho_X^{-1/4} \left( \frac{\alpha}{10^{-4}} \right)^{-3/4},
\end{equation}
where the expressions are evaluated for the default parameters of the $R=3\ \mathrm{AU}$ model (\Tb{physpar}) with ${\cal R}_\mathrm{gd} = {\cal R}_\mathrm{gc} {\cal R}_\mathrm{cd} = 100$ the gas to dust mass ratio, and where the factor of 10 in \eq{tcturb} follows from the considerations given above. For turbulence the increase in geometrical area due to the fractal growth is exactly cancelled by the decreased $\tau_\mathrm{f}$ so that the timescales for turbulence become independent of size and $\phi$ and growth progresses exponentially; for Brownian motion the growth of aggregates (in mass) is proportional to $t^2$ \citep{2004ASPC..309..369B}.

Equations\ (\ref{eq:tcBrown}) and (\ref{eq:tcturb}) show that aggregate formation is initiated by Brownian motion. Turbulence can take over at high $\alpha$ but the dust is then also quickly swept up by chondrules. At high gas densities aggregates can grow large. 

\subsubsection{Dust-chondrule/compound accretion\label{sec:dustaccr}}
The size of the dust aggregates at which they are accreted by chondrules/compounds ($a_\mathrm{dust}$) depends on the dust-chondrule accretion time $t_\mathrm{dc}$ in relation to $t_\mathrm{dd}$. The timescale, $t_\mathrm{dc}$, for a dust particle to encounter a chondrule of size $a_\mathrm{ch}$, is
\begin{equation}
  \label{eq:tcchon}
  t_\mathrm{dc} = \frac{1}{n_\mathrm{ch}\pi a^2_\mathrm{ch} \Delta v} = \frac{4 \rho_\mathrm{c}^{(s)} a_\mathrm{ch}}{3 \rho_\mathrm{c} \Delta v} = \frac{4 {\cal R}_\mathrm{gc}}{3 \mathrm{Re}^{1/4} \alpha^{1/2} \Omega} \approx 240\ \mathrm{yr}\ \rho_X^{-1/4} \left( \frac{\alpha}{10^{-4}} \right)^{-3/4},
\end{equation}
where a monodisperse distribution of chondrules that dominates the cross section ($a_\mathrm{ch} \gg a_\mathrm{dust}$) and the velocity field is assumed, and ${\cal R}_\mathrm{gc}=100$. \Eq{tcchon} again assumes the relation $\Delta v \sim v_\mathrm{s} \tau_\mathrm{ch}/t_\mathrm{s}$, although chondrules might also fall in the square-root part of the velocity regime (\se{turbmot}). Equations\ (\ref{eq:tcturb}) and (\ref{eq:tcchon}) show that in turbulence the dust is preferentially swept up by chondrules. On the other hand, in Brownian motion, velocities are always determined by the smallest particle (the dust); the grains therefore simply collide with particles that dominate the total cross-section. Thus, if turbulence dominates the velocity field for the grains and ${\cal R}_\mathrm{cd} \sim 1$, chondrules will sweep up the dust before significant aggregation takes place and $a_\mathrm{dust}\sim a_0$; contrarily, if Brownian motion dominates (or when $\tau_\mathrm{ch} \gg t_\mathrm{s}$), the monomers will first collide with each other before being accreted by chondrules.

The question that remains is what this means for the porosity of the rim. Assuming a hit-and-stick process, in which the accreting dust particles are all of equal size and much smaller than the chondrule/compound, the structure of the rim will resemble that of particle-cluster agglomeration (PCA). 
Thus, if the dust particles are solid monomers the filling factor of the rim indeed equals $\phi_\mathrm{PCA}$. On the other hand, if the accreting dust particles are fluffy aggregates, but still smaller than the chondrule, the resulting filling factor of the rim will be less than $\phi_\mathrm{PCA}$. The precise filling factor will be determined by the amount of interlocking between the aggregates but as a crude upper limit the aggregated may be approximated as a homogeneous porous sphere such that the packing process of the dust aggregates is PCA. Then, the filling factor of the porous dust, $\phi_\mathrm{pd}$, is just the product of $\phi_\mathrm{PCA}$ (caused by hit-and-stick packing) and the porosity the dust aggregates already contain ($\phi_\mathrm{dust}$), \ie\ $\phi_\mathrm{pd} = \phi_\mathrm{PCA}\phi_\mathrm{dust}$. 

In yet another collisional growth scenario we envision that chondrules are mixed into a dust region \textit{after} this dust has aggregated into dust balls but \textit{before} the dust balls are compacted \citep{2007A&A...461..215O}. In any case we assume here that the dust consists of porous aggregates and parameterize its filling factor by the $\phi_\mathrm{pd} = \phi_\mathrm{PCA}\phi_\mathrm{dust}$ relation.  Using \eq{phidust}, $a_\mathrm{dust}$ is the parameter that regulates the fluffiness of the dust accretion process and we run models at different values of $a_\mathrm{dust}$ to test its importance and sensitivity.

\subsubsection{Compound-compound accretion\label{sec:ccstick}}
The timescale for chondrule-chondrule accretion, $t_\mathrm{cc}$, is similar to the dust-chondrule timescale, \eq{tcchon}. During the collision, a fraction of the porous dust is lost to the compact dust phase. Since the porous phase filling factor is always less than that of the compact phase there is always a net loss in geometrical volume when two compounds collides, \ie\ $V < V_1 + V_2$. This, we call the `conservative approach' (no fractal accretion of compounds). Alternatively, collisions of compounds (consisting of one or more dust-rimmed chondrules) may be in the hit-and-stick regime. This would occur if the impact energy is absorbed locally and is not communicated to other parts of the compound. In that case the compound packing proceeds fractally. \citet{2007A&A...461..215O} provide an expression for the growth of $V$ in the hit-and-stick case for particles of different size, derived by an interpolation from the PCA and CCA limiting cases, \ie
\begin{equation}
  V = V_1 \left( 1 + \frac{V_2}{V_1} \right)^{\case{3}{2}\delta},
  \label{eq:fractalV}
\end{equation}
where $V_1$ is the volume of the largest of the two particles that meet and $\delta \simeq 0.95$. The growth of the porous phase $V_\mathrm{pd}$ then results from the gain in $V$ through \eq{Vgeodef}. The porous phase is then a mixture of porous dust and voids and the geometrical volume becomes a balance between hit-and-stick packing of chondrules (increasing $V$) and compaction of porous dust (decreasing $V$). This, contrary to the conservative approach in which the total volume always decreases at collision.

For example, if $V_2 = \case{1}{10}V_1$, \eq{fractalV} shows a volume $0.045V_1$ is added to the porous phase, decreasing its filling factor. It then depends on $f_\mathrm{comp}$ how the net growth of the porous phase turns out. In the initial stages of coagulation $f_\mathrm{comp}$ is often very low and, therefore, fractal accretion of compounds can become very important in enhancing the geometrical volume of the compounds.

Although fractal accretion of compounds increases the volume of the porous phase, no mass is transferred to it. It is only the filling factor that is affected, in its turn affecting the aerodynamic properties of the compound. Eventually, due to compaction, all models run out of porous dust and the dust inside the final objects -- whether fractal accretion is involved or not -- has the same filling factor $\phi_\mathrm{cd}=0.33$. The lower filling factor of the porous phase during the collision process merely reflects the voids between the chondrules that are created in the models with fractal accretion. As velocities increase, however, the fractal structure must collapse. We suspect, furthermore, that structures of very low filling factor are too weak to survive the more violent collisions \citep[see, \eg][]{2007Icar..191..779P}.

\subsection{Collisional scenario\label{sec:modsum}}
We have proposed a model where chondrules -- in the presence of dust particles -- acquire rims of fine-grained dust, which help them stick  together, and discussed various collisional scenarios for this growth process. Here we briefly summarize the envisioned scenario from a chronological viewpoint, and emphasize the (free) parameters the model contains.

At the start of our simulation ($t=0$) a population of chondrules encounters a reservoir of dust particles of characteristic size $a_\mathrm{dust}$ (see \se{dustaccr}). The physical conditions of the disk (\eg\ gas density $\rho_X$, turbulent strength $\alpha$; see \se{nebula}) determine the relative velocities between the chondrules/compounds. Another important parameter is the spatial density of dust ($\rho_\mathrm{d}$) or, rather, the dust to chondrule density ratio, ${\cal R}_\mathrm{cd}$, since this determines the thickness of the rims. These, and other physical conditions at the start of the simulation determine the subsequent accretion process. First, chondrules start to accrete the dust aggregates (see \se{dustaccr}), resulting in a porous rim of filling factor, $\phi_\mathrm{pd}$. When rimmed chondrules collide, part of this porous structure collapses to $\phi_\mathrm{cd}=0.33$ filling factor through the initiation of rolling motions. This dissipates the collisional energy and, if enough porous dust is present by the criterion of \eq{fcoll}, the two chondrules stick and a compound is created. In this way compounds are created and many chondrules can be stuck together until the accretion process is terminated when both the amount of free-floating dust and the porous dust inside the compounds have become insignificant. The end product is an inert population of compact-dust rimmed chondrules and compounds that only bounces. Collisional fragmentation is not explicitly included in the model, but we can \textit{a posteriori} compare the velocities with a critical threshold ($\sim \mathrm{m\ s^{-1}}$) to verify its importance.

\section{Monte Carlo coagulation}
\label{sec:mccoag}
The physical model of chondrule accretion contains many free (\ie\ unknown) parameters. In a statistical study of compound coagulation, we will sample these free parameters at discrete intervals such that a grid of models is created (see \se{results}). In this section we briefly describe the Monte Carlo coagulation code used to calculate the collisional evolution of the compounds.

Compounds are characterized by four numbers -- \eg\ $m$, $f_\mathrm{d}$ and $f_\mathrm{p}$ to determine the mass inside each phase, and $\phi_\mathrm{pd}$ for the filling factor of the porous phase. Therefore, a Monte Carlo code, rather than the multi-variate Smoluchowski equation, is the obvious way to solve the collisional evolution. In our code we do not keep track of the individual positions of each constituent unit (the monomers) within a compound as in \eg\ \citet{1999Icar..141..388K} but identify each compound by these four numbers. In this way we have a good characterization of the internal structure of the compound, although the precise internal structure cannot be retrieved.

The code we use is called \textit{event driven}; \ie\ the timestep $\Delta t$ defines the time between two consecutive events \citep{1975JAtS...32.1977G,1985JColloiPhys...100..116,1992JComputPhys...100..116}. Here, events are collisions between two compounds (see below for the dust). The collision rate $C_{ij}$ gives the probability of a collision -- $C_{ij}\Delta t$ is the probability of collision in the next (infinitesimal) timestep $\Delta t$ involving compounds $i$ and $j$, \ie
\begin{equation}
  C_{ij} = \frac{\sigma_{ij}\Delta v_{ij}}{ {\cal V}};\quad \mathrm{and}\quad C_\mathrm{tot} = \sum_i^N \sum_{j>i}^N C_{ij},
  \label{eq:colprob}
\end{equation}
in which $\sigma_{ij}$ is the collision cross section\footnote{In general one must distinguish between the collisional cross section (which gives the reaction rate of the two species) and the geometrical cross section (which is the average projected area of the particle that determines the coupling to the gas). Here we will simply equate them as in $\sigma_{ij} = \pi(a_i+a_j)^2$ and ignore the small discrepancy (see \citealt{2004PhRvL..93b1103K}).} between particles $i$ and $j$, $\Delta v_{ij}$ the relative velocity between the two compounds, and $N$ the total number of particles in the simulation. The volume of the simulation, ${\cal V}$, is determined from the spatial density in chondrules, the constant $\rho_\mathrm{c}$, and the total mass in chondrules, \ie\ ${\cal V} = \sum_{i=1}^N m_i(1-f_{\mathrm{d},i})/\rho_\mathrm{c}$. From these quantities the timestep is defined by $\Delta t = -C_\mathrm{tot}^{-1} \ln \overline{r}$, with $\overline{r}$ a random deviate. The particles that are involved in the collisions are also determined randomly, weighted by their collision rates $C_{ij}$. Then, using the recipes outlined in \se{model}, the outcome of the collision -- sticking or bouncing -- is determined. In either case, the parameters of the new or modified compounds are re-computed. (In the case of bouncing the change is reflected in a smaller size, $a$, due to the compaction.) Subsequently, the new collision rates of the particles (\ie\ the $\{ C_{ik} \}$ and, if the second particle due to bouncing is still present, the $\{ C_{jk}\}$ for $k = 1\dots N$ and $k\neq i,j$) are re-computed. These updates of the collision rates are the most CPU-intensive part of the code. With it one cycle is completed, after which a new stepsize $\Delta t$ is determined and the steps repeat themselves.

\Eq{colprob} involves the total relative velocity between the particles. To calculate $\Delta v_{ij}$ we use thermal, turbulent and systematic velocities (eqs.\ [\ref{eq:Brownian}], [\ref{eq:vrad}] and [\ref{eq:relvs}]), adding them up in quadrature. Strictly speaking, the zero-dimensional nature of the MC-model is inconsistent with dispersal of particles (particles do not have a positions); however, when the drift is modest the change in the physical environment is negligible and we can still use the MC-approach. For the radial drift this assumption applies only when the total drift is small compared to the initial location of the particle, \ie\ $\Delta R \ll R$, such that the same physical conditions apply throughout the simulation. We will \textit{a posteriori} check whether radial drift is significant.

Apart from collisions between compounds, we also keep track of dust accretion. This is, however, not implemented in a Monte Carlo fashion: it would have made the code very slow since the tiny dust particles far outnumber the chondrules. Instead, we catalogue the cumulative dust mass that is accreted by the compounds over the timesteps; \ie\ for compound $i$ we increase the amount by $\pi a^2_i \Delta v_{i\mathrm{d}} \rho_\mathrm{d}(t) \Delta t$. Only when this mass exceeds a certain fraction (say $f_\mathrm{upd} = 10^{-3}$) of the total mass of the compound, this quantity is added as porous dust to the compound and the $f_\mathrm{d}, f_\mathrm{p}$ parameters as well as the collision rates are updated (\eqp{colprob}). Although this procedure makes the program still a bit slow at the initial stage of the simulation, it is definitely much faster than updating all $\case{1}{2}N(N-1)$ collision rates at \textit{every} timestep. The (decreasing) amount of free-floating dust, $\rho_\mathrm{d}(t)$, is computed in this way. We have examined the sensitivity of this mechanism on $f_\mathrm{upd}$ and found that $f_\mathrm{upd}=10^{-3}$ is accurate, while much more efficient (faster) than, \eg\ $10^{-5}$.

The strong point of the MC code is that it can deal with many structural parameters and that it is transparent and straightforward; the weak point, however, is its low numerical resolution. Given the complexity of the model and the large number of models we intend to run, the number of particles ($N$) we use in the simulations is a few thousands at most. To prevent the resolution from deteriorating (a collision resulting in sticking decreases $N$ by one) we artificially stabilize the total particle number by a procedure called \textit{duplication}. In this process, one particle is randomly chosen and duplicated from the existing population \citep{1998...53..1777}. Subsequently, ${\cal V}$ is increased proportionately such that the total density in chondrules, $\rho_\mathrm{c}$, stays constant. This procedure is called the constant-$N$ algorithm -- an algorithm far superior in terms of accuracy to the constant ${\cal V}$ algorithm \citep{1998...53..1777}, and we have previously shown that it is able to calculate large orders of growth, especially when the size distribution is narrow \citep{2007A&A...461..215O}. Through the duplication mechanism, furthermore, a distinction can be made between `duplicates' and `distinct species,' and it is actually the latter that we keep constant, such that the total number of compounds involved can be much larger than a few thousands, also improving the efficiency of the model.

\section{Results}
\label{sec:results}
In our models we generally recognize three stages in the growth process: hit-and-stick dust accretion (increasing the porosity), compound accretion (\ie\ growth), and compaction with accompanied stalling of the growth. The balance between these phases controls the size of the resulting compounds, while their relative importance and `timing' are determined by the adopted model parameters. In \se{canrun} we discuss two illustrative cases, focusing on the temporal stages during their evolution. Then, in \se{parstudy} we investigate the sensitivity of the other parameters by means of a parameter study. 
\subsection{Individual model runs\label{sec:canrun}}
\begin{figure*}[t]
  \plottwo{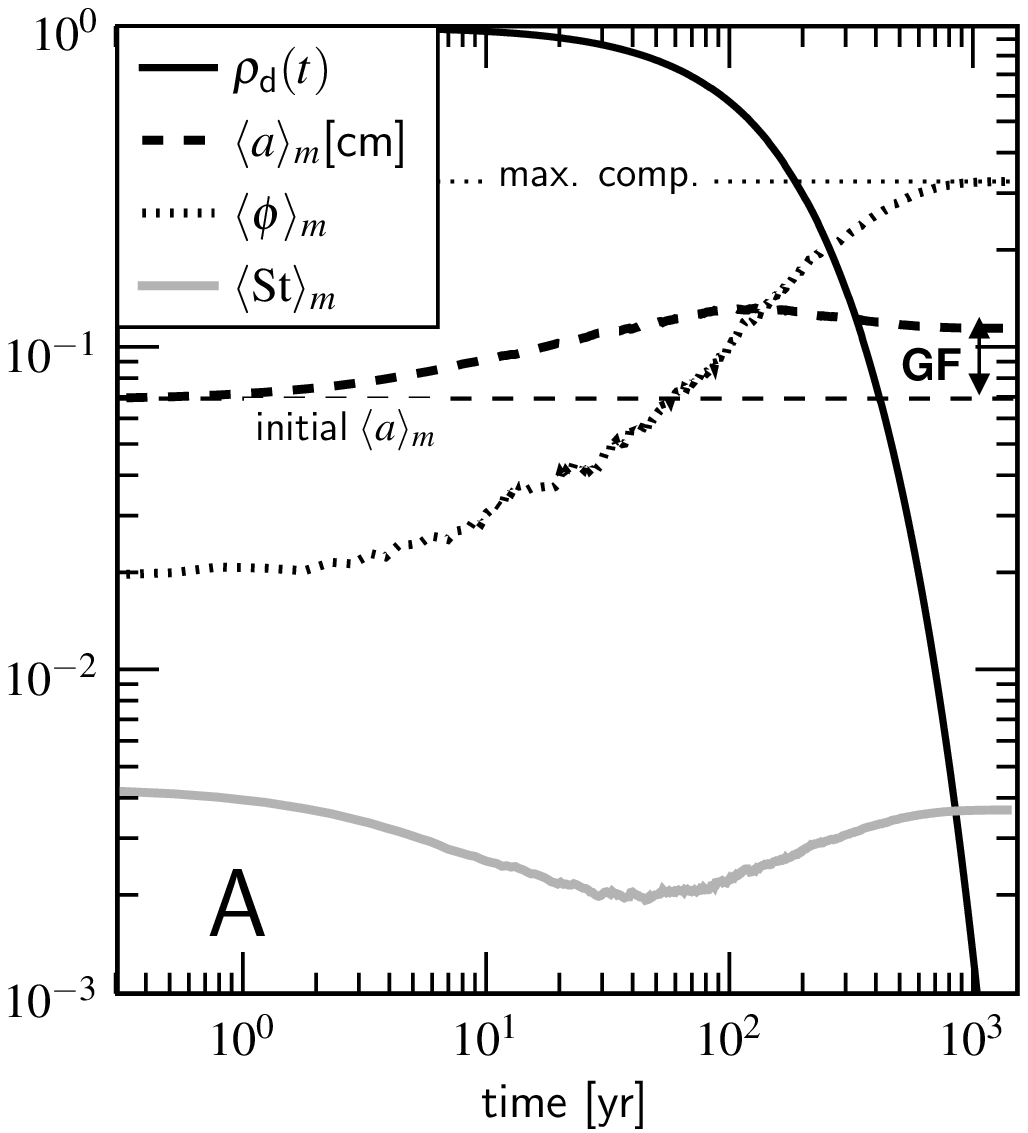}{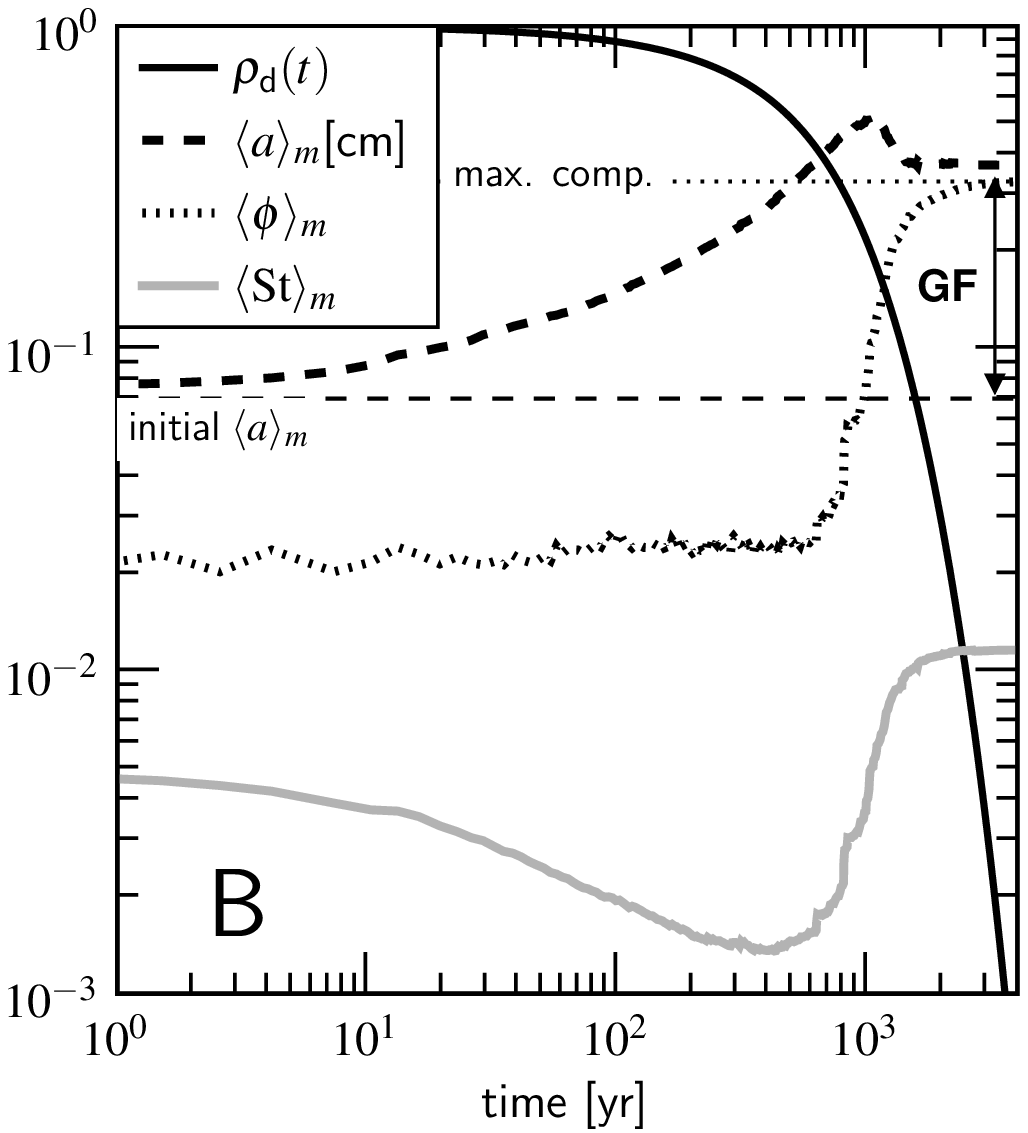}
  \caption{\label{fig:stats}(\textit{A}) A run of the compound accretion model with $\alpha=10^{-4}$, $\rho_X=1$, $a_\mathrm{dust}=10\ \mu\mathrm{m}$, $\gamma=19\ \mathrm{erg\ cm^{-2}}$, $R=3\ \mathrm{AU}$, ${\cal R}_\mathrm{gc}=100$ and ${\cal R}_\mathrm{cd}=1$ (the default model). Plotted as function of time are: the normalized density of free floating dust (solid black curve; the initial dust density is $\rho_\mathrm{dust}(t=0)=2.4\times10^{-13}\ \mathrm{g\ cm^{-3}}$); the mass-averaged size of the population (dashed-line); the mass-averaged filling factor of the dust within the compounds (dotted curve); and the mass-averaged Stokes number of the population (solid grey curve). Shown is also the definition of the growth factor, $GF$. All quantities share the same $y$-axis. (\textit{B}) Like (\textit{A}) but with $\alpha=10^{-6}$.}
\end{figure*}
\begin{figure*}[t]
  \plottwo{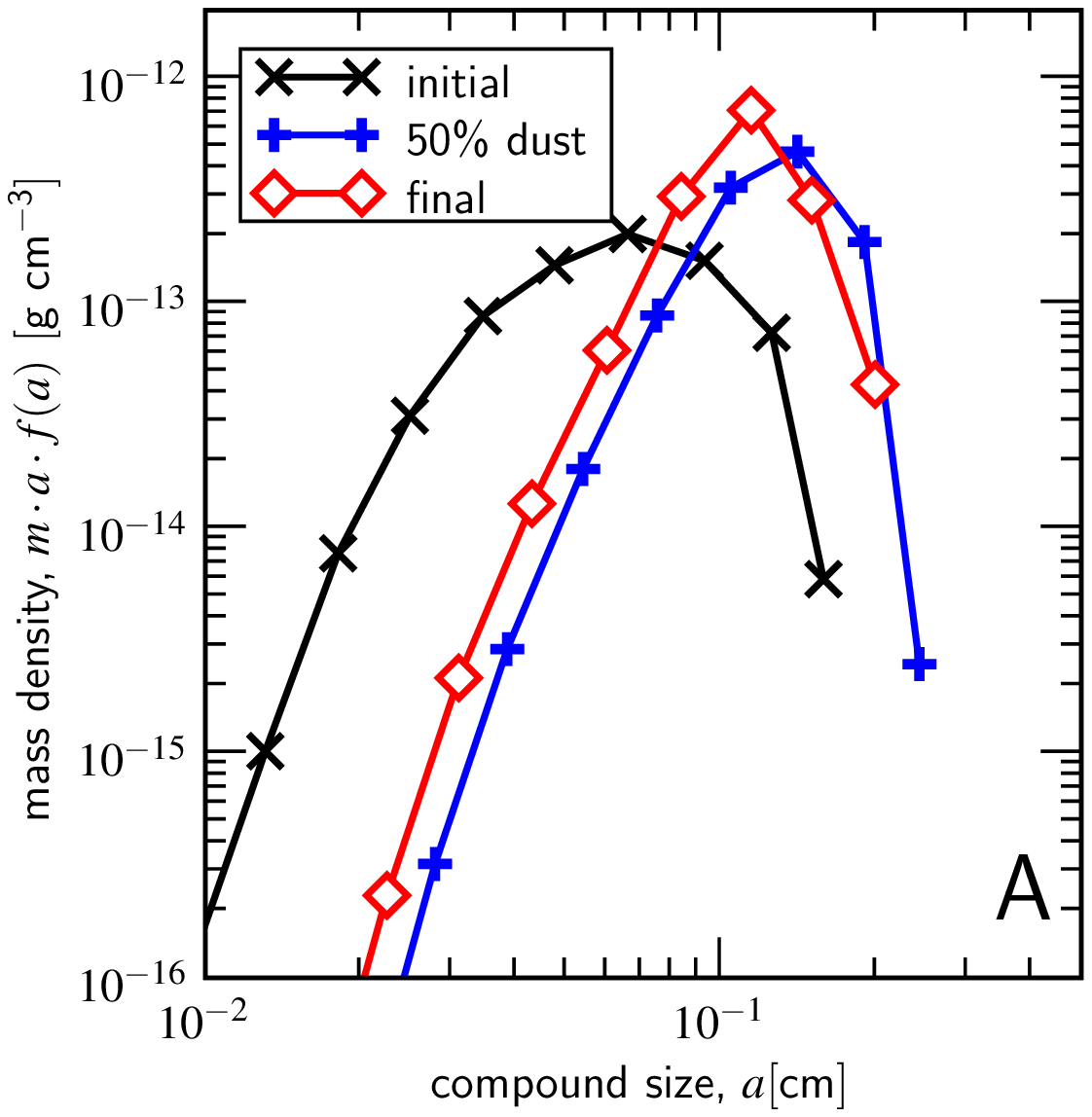}{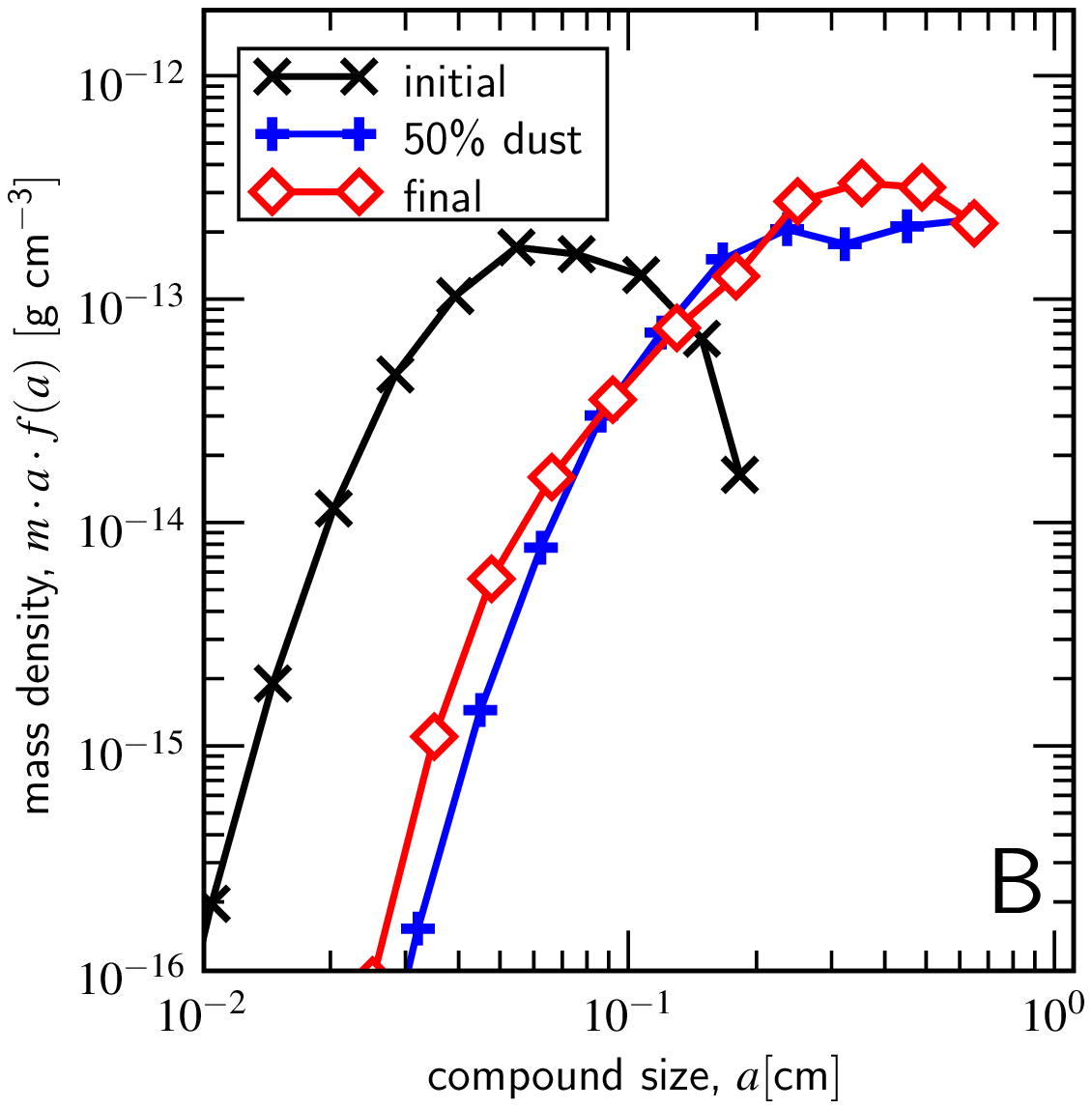}
  \caption{\label{fig:distr}Size distributions of compounds corresponding to the runs in \fg{stats} for the $\alpha = 10^{-4}$ model (\textit{A}) and the $\alpha = 10^{-6}$ model (\textit{B}). Shown are the initial distribution (crosses), the distribution at the time where 50\% of the dust has been accreted (plus-signs) and the final distribution (diamonds). Note that compaction has the effect of shifting the distribution to the left.}
\end{figure*}
Figures \ref{fig:stats} and \ref{fig:distr} show detailed results for two individual runs of the simulation with (default) parameters of gas density $\rho_X =1$, $a_\mathrm{dust}=10\ \mu\mathrm{m}$, ${\cal R}_\mathrm{gc} = 100$ and ${\cal R}_\mathrm{cd} = 1$, $\gamma = 19\ \mathrm{erg\ cm^{-2}}$ at a distance of 3 AU (see \Tb{physpar}). In these figures panels A correspond to a model with $\alpha=10^{-4}$, while $\alpha=10^{-6}$ in panels B. In \fg{stats} several (mass-averaged) quantities are shown as function of time, while in \fg{distr} the size distributions of compounds are shown at three points during their evolution: \textit{i)} $t=0$ (the initial size distribution of chondrules); \textit{ii)} the time at which 50\% of the dust is accreted; and \textit{iii)} the time at which a negligible amount of porous dust remains (the final distribution). The negligible criterion is met when both the \textit{porous} dust mass within all compounds as well as the density of free-floating dust are less than 0.1\% of the initial dust mass.

In \fg{stats} we make use of mass-weighted averages. For example, the mass-weighted average size of the population is defined as
\begin{equation}
  \langle a \rangle_m = \frac{\sum_i m_i a_i}{\sum_i m_i},
  \label{eq:massavdef}
\end{equation}
where the summation is over all particles of the simulation. It gives the mean size in which most of the mass of the population resides, and is more appropriate to describe the population than the average size, $\langle a \rangle$. In particular, adding a large number of small particles with negligible mass (density) to the population, decreases $\langle a \rangle$ but leaves $\langle a \rangle_m$ unaffected. In the following the prefix `mw-' is used as an abbreviation for `mass-weighted average of the distribution.'

\Fg{stats}A shows that dust is accreted on timescales of a few $10^2\ \mathrm{yr}$, which agrees well with previous studies \citep{2004Icar..168..484C}. This causes the size distribution (\fg{distr}) to shift towards larger sizes: the accretion of porous dust particles at low filling factor significantly increases the geometrical size of the compounds. The dotted curve in \fg{stats}A shows the mw-filling factor of the accreted dust. At the start of the simulation this equals $\phi_\mathrm{pd} = \phi_\mathrm{dust} \phi_\mathrm{PCA}$; however, collisions are energetic enough to compact the porous dust on a global scale. The decrease in friction time (solid grey curve), caused by the accretion of porous dust, therefore is only modest. (Note that even accretion of $\phi=\phi_\mathrm{cd}=33\%$ dust on chondrules would cause the friction time to decrease). Consequently, the sticking rate never increases much beyond $\sim 50\%$. After $\sim 10^2\ \mathrm{yr}$ the accretion of porous dust cannot keep pace with the compaction and sticking fails, resulting in a decrease of the compounds geometrical size (dashed curve). This results in a `retrograde motion' of the final size distribution curve in \fg{distr}A. In the $\alpha=10^{-6}$ model (\fg{stats}B) the collision velocities are much lower and, different from the $\alpha=10^{-4}$ model, the porous dust does not experience compaction for a long time. The sticking rate then increases to almost 100\%. However, depletion of dust triggers the end of the growth phase; growth is quickly terminated by the mutually enforcing processes of rim compaction and increasing velocities. From these panels it is clear that much growth can be achieved when relative velocities are kept low during the dust and chondrule accretion.

\begin{figure}[tbp]
  \plotone{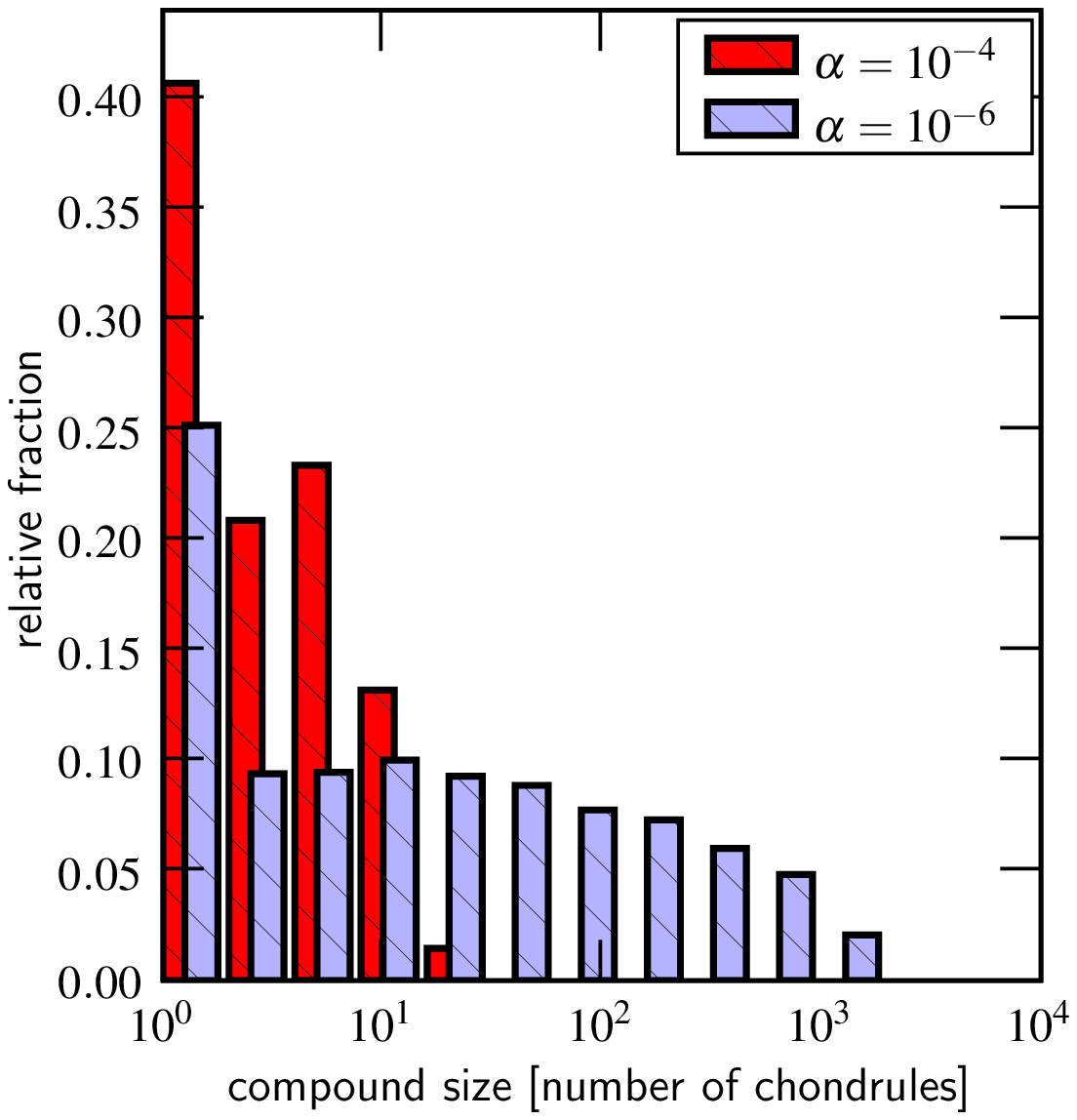}
  \caption{\label{fig:hist}Relative fraction of dust accreted by compounds of different size for the $\alpha=10^{-4}$ and $\alpha=10^{-6}$ simulations. The compounds are placed in bins according to the number of chondrules they contain. The bins are exponentially distributed by factors of two. The histogram shows the distribution of the dust over the compound sizes  (in terms of number of chondrules inside the compound) at the time of the dust accretion. Single chondrules (first bin) accrete a significant fraction of the dust.}
\end{figure}
Although the Monte Carlo code does not keep track of the position or size distribution of chondrules within compounds, we can still extract useful statistical information from the model runs. One such statistic is the distribution of the dust over compounds of different size: is the dust primarily accreted by individual chondrules or by large compounds containing many chondrules? The results are presented in the histogram of \fg{hist}. The $x-$axis denotes the number of chondrules a compound contains ($N$) and is divided into logarithmic bins of base 2, \ie\ the first bin corresponds to single-chondrule compounds, the second to compounds that contain 2 or 3 chondrules, the third to $4-7$ chondrules, etc. The $y-$axis gives the relative fraction of the dust that first accretes onto a chondrule or compound with size in each bin; that is, \fg{hist} only reflects the dust accretion history and does not include the subsequent re-distribution of dust due to coalescence of compounds (which would shift the dust-rimmed chondrules to a larger compound bin). The relatively high level of the first bin (single-chondrule compounds) reflects dust accretion during the early phase of the simulation where individual chondrules provide a high surface area and the density of free-floating dust is highest. In simulations with strong turbulence this fraction becomes very high: the dust is then accreted by single chondrules only. But even in the case of low $\alpha$ single chondrules are responsible for a significant share of the dust sweep-up, as the $\alpha=10^{-6}$ results show. Besides, larger compounds also have a larger surface to spread this dust over; rims created by dust accretion are therefore thickest on chondrules.

\subsection{Parameter study\label{sec:parstudy}}
\begin{deluxetable}{lllll}
  \tablecaption{\label{tab:freepar} free model parameters}
  \tablehead{ \colhead{(1)} & \colhead{(2)} & \colhead{(3)} & \colhead{(4)} & \colhead{(5)} }
  \startdata
   turbulent strength             & $\alpha$                                          &                       &5 & [$10^{-7}$|$10^{-3}$]               \\
   size of dust aggregates        & $a_\mathrm{dust}$                                 & cm                    &3 & $[10^{-4}$|$10^{-2}]$\\
   gas density\tablenotemark{a}   & $\rho_X$                                          & $\mathrm{g\ cm^{-3}}$&3 & [1|100]   \\
   nebula location                & $R$                                               & AU                    &3 & [1, 3, 10]   \\
   gas-chondrule ratio            & ${\cal R}_\mathrm{gc}$                            &                &2 & [10, 100]                            \\
   chondrule-dust ratio           & ${\cal R}_\mathrm{cd}$                            &                &2 & [1, 10]                              \\
   dust composition\tablenotemark{b}         & $\gamma$                                          & $\mathrm{ergs\ cm^{-2}}$&2& [19, 370]   \\
   compaction mode\tablenotemark{c}          & $f_\mathrm{geo}$                                  &                       &2 & [X, 1]       \\
   \multicolumn{2}{l}{fractal accretion of compounds\tablenotemark{d}}                &          &2         & [yes, no] 
  \enddata
    \tablenotetext{a}{$\rho_X$: gas density over MSN model at 3 AU.}
    \tablenotetext{b}{$\gamma$: energy surface density. The high $\gamma$ model corresponds to ice as the sticking agent (10 AU models only).}
    \tablenotetext{c}{$f_\mathrm{geo}= \mathrm{'X'}$: $f_\mathrm{geo}$ is computed after the procedure outlined in \se{dissip}; $f_\mathrm{geo}=1$: use $f_\mathrm{geo}=1$ always.}
    \tablenotetext{d}{Whether a hit-and-stick packing model for compounds (leading to a fractal structure, \se{ccstick}) is adopted or not.}
  \tablecomments{\strut List of free model parameters. Columns denote: (1) parameter description; (2) symbol; (3) unit; (4) number of grid points per parameter; (5) parameter range, with a grid point at every factor of 10, unless otherwise indicated. See also Table\ \ref{tab:physpar} for other (fixed) parameters.}
\end{deluxetable}
\begin{figure*}[p]
  \includegraphics[width=0.8\textwidth]{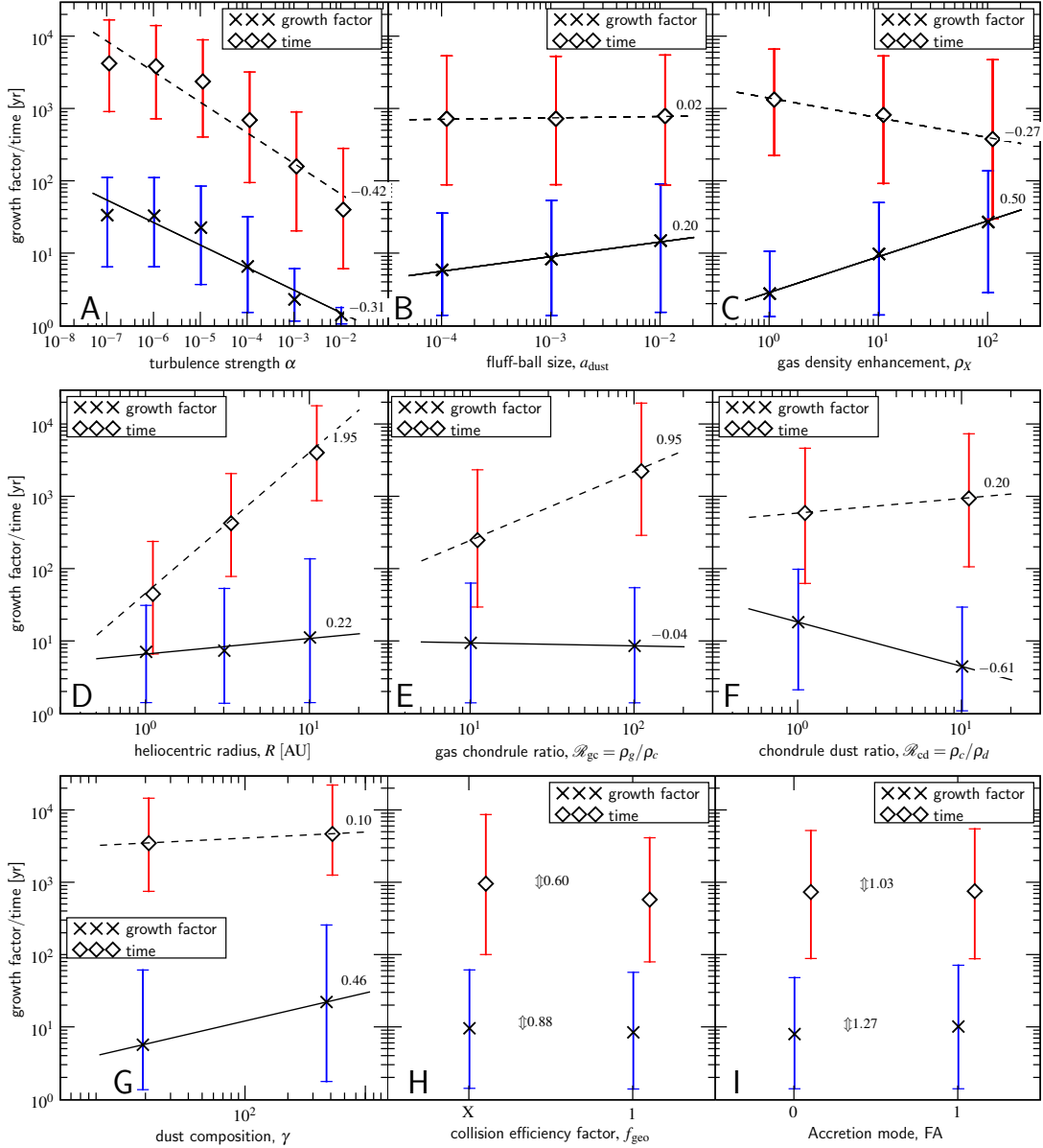}
  \caption{\label{fig:trends}Results of the parameter study. Each panel sorts the data according to a free parameter ($x$-axis), from which the logarithmic mean and variance are calculated. Two output values are shown on the same $y$-axis: growth factor (crosses) and simulation time (diamonds). The lines show the trend in variation of the parameter and the `best-fit' power-law exponent is given. (\textit{G}) Data from $R=10\ \mathrm{AU}$ models only, comparing silicate dust ($\gamma = 19\ \mathrm{erg\ cm^{-2}}$) with ice ($\gamma=370\ \mathrm{erg\ cm^{-2}}$). (\textit{H}) $f_\mathrm{geo}=$ X indicates $f_\mathrm{geo}$ is a free parameter calculated after \eq{intersect}, while $f_\mathrm{geo}=1$ indicates it is 1 always. (\textit{I}) 0 and 1 denote, respectively, that fractal accretion of compounds is turned off or on. In these latter two panels the numerical factor next to the $\Updownarrow$ gives the ratio in growth factor and timescale between the two modes (not the power-law exponent). }
\end{figure*}
\Fg{trends} presents the results of the parameter study. The free parameters (Table\ \ref{tab:freepar}) are distributed over a grid such that each grid point corresponds to a unique model. In total a few thousand distinct models are run. Each model is run a few times to account for stochastic effects in their results (typically $\sim$10\% or so). For each free parameter the models are ordered by the grid-values of the parameter, corresponding to the panels in \fg{trends}. Two output values are shown: the ratio of final mw-radius to initial mw-radius, or growth factor (crosses) and the time at which the dust is depleted and the simulation terminated, or simulation time (diamonds) (they share the same $y$-axis; for clarity the timescale error-bars are slightly offset in the $x$-direction). The symbols denote the logarithmic averages of all models at the grid-values and the error bars indicate the range in which 50\% of the models fall. This spread can be huge since it is primarily determined by the spread in the other parameters (and therefore nowhere close to Gaussian). The same holds for the averages: these can be arbitrarily scaled up or down by giving more weight to extreme models in the parameter study.

However, the value of \fg{trends} lies not in its absolute numbers but in the trends that emerge from the parameter variation. The lines indicate this trend and their slopes are given in each panel. From these, it is seen that timescales are primarily determined by turbulent $\alpha$ (velocities), nebula location $R$ (densities) and the chondrule density (panel E). Parameters that favor large growth of compounds are low $\alpha$ (panel A), high gas densities (panel C), low chondrule-to-dust ratios (panel F), and high surface energy densities (panel G). Growth is favored in these models due to the moderate relative velocities (panels A and C) or better sticking capabilities (panels F and G). Other parameters are sometimes surprisingly irrelevant. For example, the dependence on the size of the dust fluff-balls, $a_\mathrm{dust}$, defining their porosity (\eqp{phidust}), is only modest (panel B), and also the latter two panels do not show clear trends. Panels H and I directly give the ratio between the two data points, instead of the exponent of the power-law fit. Panel H shows the effects of taking the geometry of the collision into account. $f_\mathrm{geo} = \mathrm{X}$ where $0 < \mathrm{X} \le 1$ means that $f_\mathrm{geo}$ is determined by the geometry of the collision as discussed in \se{dissip}, whereas $f_\mathrm{geo}=1$ indicates that all dust is available for compaction. However, allowing for a lower $f_\mathrm{geo}$ also reduces the maximum amount of dust that is compacted. Apparently, these two effects largely compensate. A similar insensitiveness is shown in panel I: whether we allow for fractal accretion of compounds (`1') or not (`0') does not, on average, make a difference. Note, however, that the values shown in the panels are averages; in some individual simulations we do see a notable increase when fractal accretion is turned on.

\begin{figure*}
  \plotone{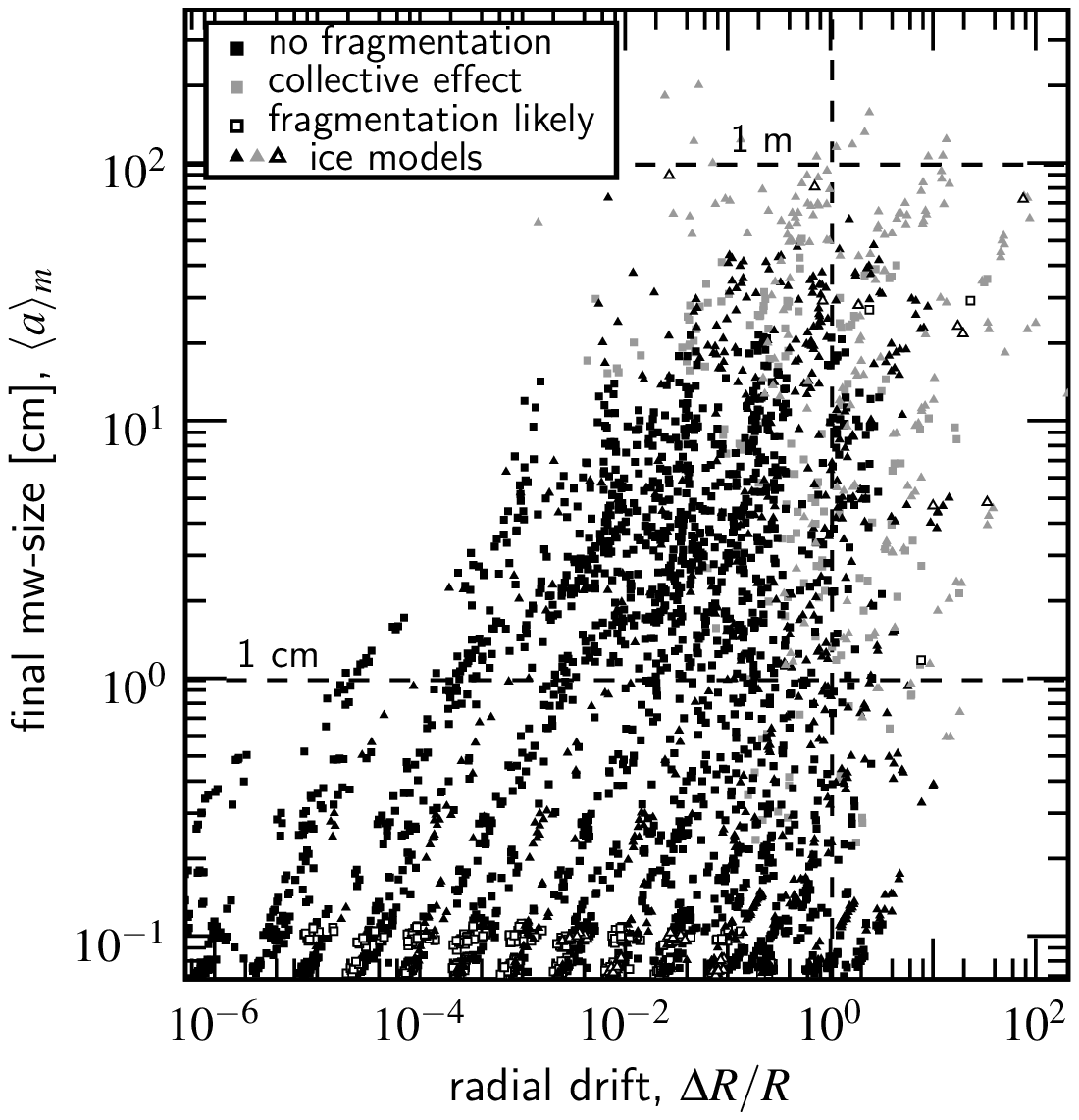}
  \caption{\label{fig:scatter}Scatter plot of the fractional inward radial drift covered during the aggregation process ($\Delta R/R$, $x$-axis) against the final mass-averaged size of the compounds ($\langle a \rangle_m$, $y$-axis). The results of all model runs are shown. Models are separated into the low-velocity regime ($\Delta v \lesssim 10^2\ \mathrm{cm\ s^{-1}}$, black squares) and the high velocity regime ($\Delta v \gtrsim 10^2\ \mathrm{cm\ s^{-1}}$, open squares). Triangles denote `ice models' ($\gamma=370$ at $R=10\ \mathrm{AU}$). In the models shown by grey squares (or triangles) the collective effect could have prevented high drift velocities but this is not incorporated in the present models (see text). The vertical dashed line corresponds to a drift of 1 AU. The dashed horizontal lines indicate compound sizes of $\mathrm{1\ cm}$ and $\mathrm{1\ m}$, respectively.}
\end{figure*}
Panel A shows that the positive correlation between growth factor and decreasing $\alpha$ breaks down for values below $\alpha < 10^{-5}$: the growth flattens out and reaches a constant level. The reason is that for low values of $\alpha$, and, subsequently, large compounds, radial drift motions quickly take over from turbulent motions such that the evolution becomes insensitive to $\alpha$. The high radial drift obtained when particles approach the $\mathrm{St}=1$ regime forms a barrier for further growth. Low $\alpha$ combined with high gas densities delay this transition since compounds are now much better coupled to the gas (lower Stokes numbers) meaning much growth early on. Yet, Stokes numbers inevitably grow to values near unity, and in most cases the resulting $\eta v_\mathrm{K}$ drift velocities (\eqp{vrad}) stall growth below 1 meter.

Panel D shows that growth depends only modestly on nebula radius, $R$. Here, the positive correlation with growth factor is a bias resulting from the high $\gamma$ `ice models' (panel G) -- ice promotes sticking -- that are only present at $R=10\ \mathrm{AU}$. Thus, despite the fact that almost all nebula parameters scale with $R$, their combined effect does not result in a clear trend that favors growth. For example, larger nebula radii mean lower densities and higher Stokes number, increasing the velocity in the initial stages, but this is offset by a (slightly) lower sound speed, and the better sticking agents that are available.

In \fg{scatter} all models are combined in a scatter plot of total (mass-weighted) radial drift against the final mw-size obtained in the simulation. The few models that cluster around the meter size are all ice models ($\gamma=370$ and $R=10\ \mathrm{AU}$, indicated by triangles). Some of them do manage to cross the $\langle \mathrm{St}\rangle_m=1$ barrier (at 10 AU and $\rho_X=1$ this already happens at a few centimeters) but do not make the jump to planetesimal sizes. In models that during their growth drift less than $\Delta R \sim R$ the local assumption is justified; for models that drift over several AU-distances, however, the approximation we used in the calculation of the collisional evolution, \ie\ that the physical conditions stay the same, breaks down. Note, however, that the drift distances in \fg{scatter} are upper limits: radial drift slows down with decreasing $R$ due to a better coupling to the gas, or may diminish when collective effects become important (see \se{colleff}).

\subsection{Importance of fragmentation}
The models neglect the possibility that high velocity collisions will fragment, rather than merely compact or bounce, compounds. In \se{impfrag} it was estimated that at $\sim \mathrm{m\ s^{-1}}$ velocities, fragmentation becomes likely, starting with erosion, followed by catastrophic disruption of the compound. This threshold can now be compared to the maximum velocities attained at a particular Stokes number (see \fg{motions}), \ie\ $\sim \eta^{1/2} c_\mathrm{g} \mathrm{St}$ and $\sim \alpha^{1/2} c_\mathrm{g} \mathrm{St}^{1/2}$ for the systematic drift and turbulent velocities, respectively. Inserting the final mw-Stokes number into these expressions, we obtain a criterion whether fragmentation is of importance. As the critical velocities we take $2\ \mathrm{m\ s^{-1}}$ for turbulence and $6\ \mathrm{m\ s^{-1}}$ for radial drift (see \fg{fcomp}). It can be shown that for systematic velocities scaling proportional to size the collisional energy peaks at size ratios of $\epsilon \sim 0.5$, which, according to \fg{fcomp}, corresponds to a fragmentation velocity of $\sim 3\ \mathrm{m\ s^{-1}}$, or a $\sim 6\ \mathrm{m\ s^{-1}}$ radial drift velocity for the largest particle. For ice models (triangles) the fragmentation threshold is increased by another factor of four, reflecting their higher $\gamma$. In \fg{scatter} the models that have crossed the threshold velocity are indicated by an open square, whereas black squares indicate velocities that stay below the threshold. The grey squares are models in which collective effects could have had a significant reduction in relative velocities and drift rates, due to concentration of compounds near the midplane (see below, \se{colleff}). However, these subtleties are presently \textit{not} taken into account in the simulation and it remains unclear whether fragmentation is an important phenomenon in models for which settling is important.

\begin{figure}
  \plotone{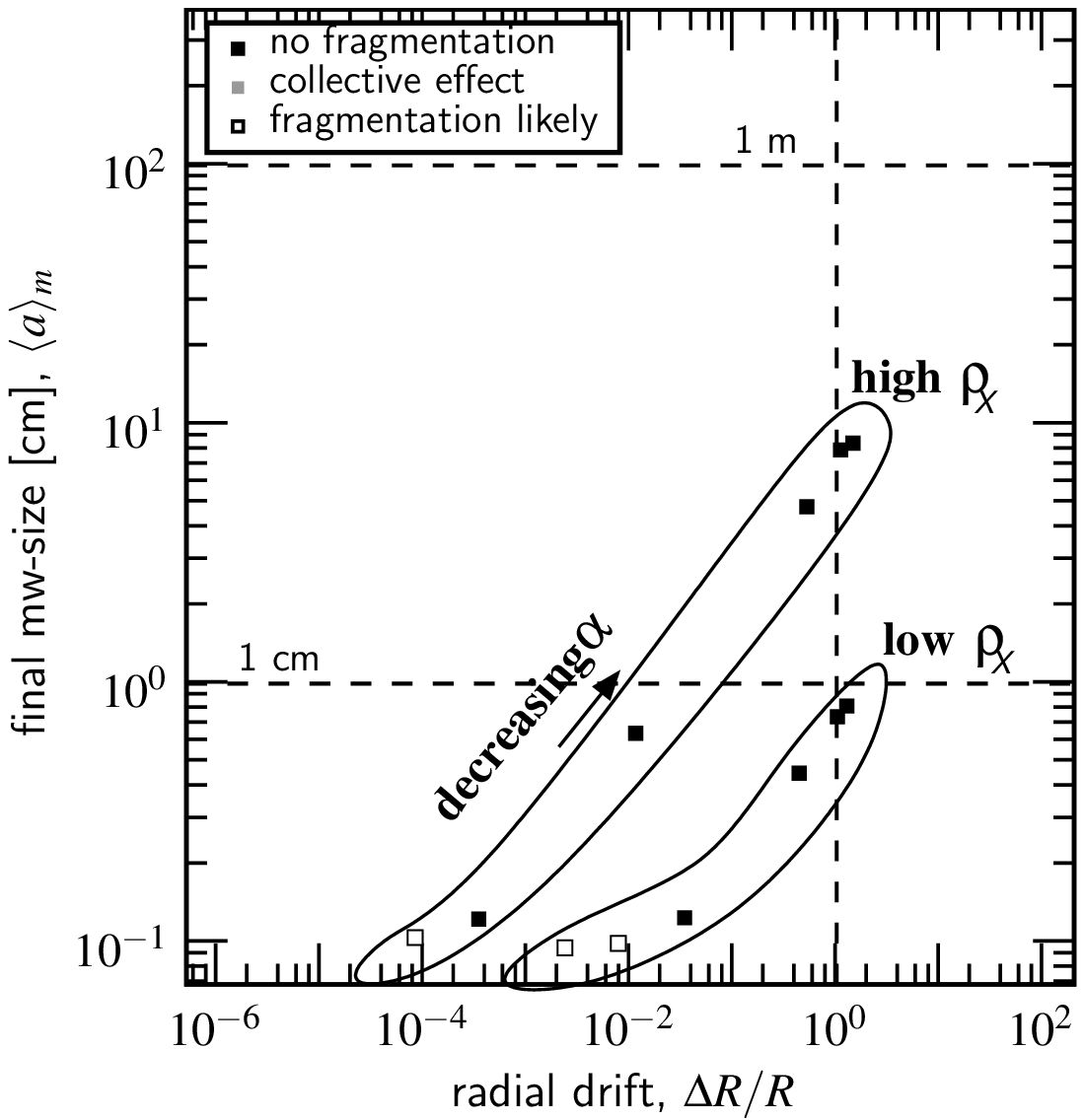}
  \caption{\label{fig:scatter2}A selection of 12 models from the scatter plot of \fg{scatter}, indicating systematic trends. Compared to \fg{scatter} results are limited to: $a_\mathrm{dust}=10^{-3}$, $\rho_X=1$ or 10, ${\cal R}_\mathrm{gd}=100, {\cal R}_\mathrm{cd}=1, R=\mathrm{3 AU}$, $f_\mathrm{geo}=\mathrm{X}$ (local compaction), and no fractal accretion of compounds. See \Tb{selectmod} for quantitative results.}
\end{figure}
\begin{deluxetable}{rrrrrrrrr}
  \tablecaption{\label{tab:selectmod}Detailed results}
  \tablehead{ \colhead{$\alpha$} & \colhead{$\rho_X$}  & \colhead{$\langle a \rangle_m$} & \colhead{$\langle \mathrm{St} \rangle_m$} & \colhead{$\rho_\mathrm{p}/\rho_g^\mathrm{mid}$}  & \colhead{$\Delta v^\mathrm{turb}$}       &  \colhead{$\Delta v^\mathrm{sys}$}       &  \colhead{$\Delta v^\mathrm{sys, CE}$} \\
            &           & [cm]                  &                                 &                               & [$\mathrm{cm\ s^{-1}}$] & [$\mathrm{cm\ s^{-1}}$]  &  [$\mathrm{cm\ s^{-1}}$]  \\
  (1)       & (2)       & (3)               & (4)                   & (5)                             & (6)                           & (7)                 & (8)                     }
  \startdata
{$10^{-7}$} & {$10$} & {$8.5$} & {$2.6\times10^{-2}$} & {$4.9$} & {$5.9$} & {$358.6$} & {$10.3$} \\
{$10^{-6}$} & {$10$} & {$8.0$} & {$2.5\times10^{-2}$} & {$1.5$} & {$18.0$} & {$336.9$} & {$54.6$} \\
{$10^{-5}$} & {$10$} & {$4.8$} & {$1.5\times10^{-2}$} & {$0.4$} & {$44.3$} & {$204.4$} & {$108.2$} \\
{$10^{-7}$} & {$1$} & {$0.8$} & {$2.5\times10^{-2}$} & {$4.8$} & {$5.8$} & {$345.0$} & {$10.2$} \\
{$10^{-6}$} & {$1$} & {$0.7$} & {$2.3\times10^{-2}$} & {$1.4$} & {$17.4$} & {$316.7$} & {$53.4$} \\
{$10^{-4}$} & {$10$} & {$0.6$} & {$2.0\times10^{-3}$} & {$0.0$} & {$51.2$} & {$27.4$} & {$25.1$} \\
{$10^{-5}$} & {$1$} & {$0.5$} & {$1.4\times10^{-2}$} & {$0.4$} & {$42.8$} & {$190.9$} & {$103.8$} \\
{$10^{-3}$} & {$10$} & {$0.1$} & {$3.9\times10^{-4}$} & {$0.0$} & {$71.4$} & {$5.3$} & {$5.2$} \\
{$10^{-4}$} & {$1$} & {$0.1$} & {$4.0\times10^{-3}$} & {$0.1$} & {$71.9$} & {$53.9$} & {$47.9$} \\
{$10^{-3}$} & {$1$} & {$0.1$} & {$3.3\times10^{-3}$} & {$0.0$} & {$206.7$} & {$44.5$} & {$42.8$} \\
{$10^{-2}$} & {$1$} & {$0.1$} & {$3.2\times10^{-3}$} & {$0.0$} & {$641.7$} & {$42.9$} & {$42.0$} \\
{$10^{-2}$} & {$10$} & {$0.1$} & {$3.5\times10^{-4}$} & {$0.0$} & {$212.3$} & {$4.7$} & {$4.6$} 
  \enddata
  \tablecomments{Detailed results from 12 selected models (see text), ordered after final mw-size, Col.\ (3). The columns denote: (1) turbulent-$\alpha$; (2) gas-density enhancement (restricted to 1 or 10); (3) final mw-size; (4) final mw-Stokes number; (5) final midplane dust-gas density ratio would settling have been included; (6) turbulent velocity contribution, $\alpha^{1/2} \mathrm{St}^{1/2} c_\mathrm{g}$; (7) systematic drift after \eq{vrad}; (8) systematic drift due to collective effects after \eq{vcolldrift} with Col. (4) for $\mathrm{St}$.}
\end{deluxetable}

For most models in \fg{scatter} fragmentation is not a serious concern. This is a natural result as compaction precedes fragmentation and growth stalls before reaching the fragmentation threshold. However, for large, fluffy compounds the compaction is more pronounced, resulting in a significant decrease in surface area-to-mass ratio, increasing the Stokes, and thereby possibly breaching the threshold for fragmentation. Also note the fragmentation models (open symbols) at the bottom of \fg{scatter}: in these the fragmentation threshold was already exceeded at the start of the simulation.

\Fg{scatter2} shows a small subset of models from \fg{scatter} that takes away the redundancy (caused by less influential parameters) and focuses on the more plausible scenarios. More specifically, \fg{scatter2} shows models limited to the following parameters: ${\cal R}_\mathrm{gc} = 100$, ${\cal R}_\mathrm{cd}=1.0$, $a_\mathrm{dust} = 10^{-3}\ \mathrm{cm}$, $\rho_X = $ 1 or 10, $R=\mathrm{3\ AU}$, $\gamma = 19\ \mathrm{erg\ cm^{-2}}$; furthermore, we assume only local compaction ($f_\mathrm{geo}=\mathrm{X}$) and assume collisions between compounds are not in the hit-and-stick regime (\se{ccstick}). Only 12 models are then shown with the only free parameters being $\alpha$ (all 6 distinct values) and $\rho_X$ (2 values). Table \ref{tab:selectmod} shows various output values corresponding to the `top ten' models of \fg{scatter2}, ordered after final mw-size; for example, the maximum velocities due to systematic and turbulent motions. This shows that for these low-$\alpha$ models systematic drift velocities (Col.\ (7)) quickly become dominant over turbulent motions (Col.\ (8)).

\section{Discussion}
\label{sec:discuss}
\subsection{Collective effects in a settled layer\label{sec:colleff}}
Despite the ability of the chondrule-sticking model to tweak many parameters to optimize the growth, compounds never achieve planetesimal sizes. Ultimately, m/s or higher velocities are unavoidable in any model due to the radial drift; that is, compounds inevitably reach (and have to cross) the $\mathrm{St}=1$ bottleneck at which their radial drift velocities peak. The studied accretion mechanism -- chondrule sticking by compaction of initially fluffy dust -- is simply too weak to grow past the $\mathrm{St}=1$ bottleneck.  

There is one issue, however, that is unaccounted for within the framework in which the simulations are performed: if the turbulence is weak enough, in addition to moving radially, compounds can also settle into a dense layer at the midplane as their Stokes numbers increase. When the density of solids at the midplane exceeds the gas density, the gas is dragged with the particles (instead of the other way around), resulting in gas velocities that tend to become closer to Keplerian, which subsequently diminishes the radial drift and relative velocities of particles. \citet{1986Icar...67..375N} have solved the equations of motion in such a two-fluid medium analytically for a single particle size (or Stokes number); the radial drift velocity now becomes (instead of \eqp{vrad}) 
\begin{equation}
  v_\mathrm{r} = \frac{2\mathrm{St}}{\mathrm{St}^2 + (1+\rho_\mathrm{p}(z)/\rho_\mathrm{g})^2} \eta v_\mathrm{K},
  \label{eq:vcolldrift}
\end{equation}
where $\rho_\mathrm{p}(z)$ is the total density of particles at a height $z$ above the midplane. For a generalized solution over a particle size distribution see \citet{1997Icar..127..290W} or \citet{2005ApJ...625..414T}. Thus, in a dust-dominated layer the radial drift of individual particles depends through $\rho_\mathrm{p}(z)$ on the density of all other particles: a collective effect. The particle concentration can be found by balancing the gravitationally induced settling rate with the diffusion rate of a particle, assuming a steady-state distribution. The scaleheight of the resulting particle distribution, $h_\mathrm{p}$, can be calculated as \citep{1995Icar..114..237D}
\begin{equation}
  h_\mathrm{p} =  \frac{H_\mathrm{g}}{\sqrt{1 + {\cal S}}},
  \label{eq:scaleheight}
\end{equation}
where ${\cal S} = \mathrm{St}/\alpha$. Under conditions of initial cosmic abundances, in order to reach $\rho_\mathrm{p}\sim\rho_\mathrm{g}$ the particles must settle into a layer of thickness one-hundredth of the gas scaleheight, requiring ${\cal S} > 10^4$ \citep{2005ASPC..341..732C}. This may occur for chondrules in very low-$\alpha$ environments, or, at moderate $\alpha$, only for large compounds during their growth and settling stage. In \Tb{selectmod} we have calculated the density enhancement ($\rho_\mathrm{p}/\rho_\mathrm{g}$, Col.\ (5)) and the corresponding velocities ($\Delta v^\mathrm{sys, CE}$, Col.\ (8)) for a few selected models at the end of their simulation, where we fixed most parameters at their default 3 AU values, except for $\alpha$ and $\rho_X$. Note that in the context of our current model setup collective effects are purely hypothetical (we treat $\rho_g/\rho_c = {\cal R}_\mathrm{gc}$ as a constant); the columns of \Tb{selectmod} therefore merely provide an indication of what could be expected had settling-effects been included. In these calculations we have used the mass-averaged Stokes number of the population (Col.\ (4) of \Tb{selectmod}) as the Stokes number that enters equations\ (\ref{eq:vcolldrift}) and (\ref{eq:scaleheight}). The last two columns of \Tb{selectmod} show that collective effects ($\rho_\mathrm{p}/\rho_\mathrm{g} > 1$) quickly reduce the radial drift. In a future study, we intend to investigate the effects of the particle concentration on the compounds' growth.

There is yet another subtlety involved when collective effects (\ie\ a dust-dominated midplane) become important. This is the Kelvin-Helmholtz instability \citep{1980Icar...44..172W}, caused by the shear between the two fluids now moving at a relative velocity of $\Delta V$, the azimuthal velocity difference between the gas in the particle-dominated and the gas-dominated layer. For shear turbulence the turbulent viscosity is $\nu_T \sim  (\Delta V)^2/\Omega \mathrm{Re^*}^2$ \citep{1993Icar..106..102C}, where $\mathrm{Re}^\ast$ is a critical Reynolds number at which the flow starts to become turbulent, which \citet{1999JGR...10430805D} found to be $\mathrm{Re}^\ast \sim 20-30$. Also, the large eddy turnover frequency in shear turbulence ($\Omega_\mathrm{e}$) can become much larger than $\Omega$, depending on the thickness of the shear layer (see \citealt{2006Icar..181..572W} for how $\Omega_\mathrm{e}$ depends on the particle density structure, $\rho_\mathrm{p}(z)$). Equating $\nu_\mathrm{T} \sim  (\Delta V)^2/\Omega \mathrm{Re^*}^2$ with $(v^\mathrm{shear}_\mathrm{L})^2/\Omega_\mathrm{e}$ then provides the expression for the shear turbulent (large eddy) velocity, $v_\mathrm{L}^\mathrm{shear}$,
\begin{equation}
  v_\mathrm{L}^\mathrm{shear} \sim \left( \frac{\Omega_\mathrm{e}}{\Omega} \right)^{1/2} \frac{\Delta V}{\mathrm{Re^*}} \approx 0.033 \left( \frac{\Omega_\mathrm{e}}{\Omega} \right)^{1/2} \Delta V \lesssim 0.18 \eta^{1/2} c_\mathrm{g},
\end{equation}
where the upper limit assumes $(\Omega_\mathrm{e}/\Omega) \sim \mathrm{Re}^\ast = 30$ and $\Delta V = \eta v_\mathrm{k} = \eta^{1/2} c_\mathrm{g}$. This corresponds to the situation where the shear layer is thin (meter-size or larger particles; the shear layer cannot become thinner than the Eckman layer, see \citealt{1993Icar..106..102C}). In that case, setting $v_\mathrm{L}^\mathrm{shear} = \alpha_\mathrm{shear}^{1/2} c_\mathrm{g}$ the equivalent $\alpha$ value for shear turbulence becomes $\alpha_\mathrm{shear} \sim 3\times 10^{-5}$. This is an upper limit; for smaller particles, or a size-distribution of particles, both $\Omega_\mathrm{e}$ and $\Delta V$ are lower and $\alpha_\mathrm{shear}$ decreases as well. Shear turbulence may therefore be much more conducive to compound growth.

Future studies must show whether these effects enable growth to planetesimal sizes. Recently, \citet{2007Natur.448.1022J} have suggested that concentration of meter-size particles ($\mathrm{St}\sim1$) in certain azimuthally-oriented near-midplane high pressure zones, which form between large turbulent eddies, might lead to gravitationally bound clumps with the mass of planetesimal size objects. The results from \citet{2007Natur.448.1022J} were most pronounced when the turbulent intensities were moderately high (this leads to the largest radial pressure contrast), suggesting values of $\alpha \sim 10^{-3}$. However, our results suggest that it is difficult to grow a population of meter-size boulders in the first place under such conditions. The maximum growth (in terms of Stokes number) our models achieve for $\alpha=10^{-3}$ is $ \mathrm{St} \sim 5 \times 10^{-3}$ at 3 AU (essentially no growth at all: just dust-rimmed chondrules). 10 AU ice models do somewhat better: $\mathrm{St} \sim7.4 \times 10^{-2}$. Even if they can form, a population of meter-sized boulders may be difficult to maintain if these originated from dust-coated, solid chondrules as modeled in this paper. In \se{impfrag} we estimated that fragmentation occurred at a critical velocity of $\sim 10^2\ \mathrm{cm\ s^{-1}}$, 30 times smaller than the expected value of $\mathrm{St}=1$ particles for $\alpha = 10^{-3}$. This translates into a specific kinetic energy for disruption of $Q^*=10^4\ \mathrm{erg\ g^{-1}}$, much lower than the critical $Q^*$ \citet{2007Natur.448.1022J} adopt (for aggregates of solid basalt objects, as taken from \citealt{2000SSRv...92..279B}). Thus, our results indicate that it may be difficult for the instability described by \citet{2007Natur.448.1022J} to become viable in the turbulent inner (ice-free) nebula.

In the outer solar system, however, conditions may be more favorable to growth in a turbulent environment. First, if ice acts as the sticking agent $Q^*$ may be over an order of magnitude larger, reflecting the scaling with the surface energy density parameter, $\gamma$. Second, if chondrule formation is not common the particles grow directly from aggregates of tiny grains to larger aggregates and therefore contain roughly twice as much mass in small grains as dust-rimmed solid chondrules.  Moreover, Stokes numbers for the same particles increase with larger heliocentric radii ($R$) due to the lower gas densities. Therefore, at large $R$ the $\mathrm{St}\sim1$ regime is reached at smaller sizes (centimeters), which may be somewhat more difficult to disrupt (\ie\ higher $Q^*$) than m-size bodies \citep{1990Icar...84..226H,2000SSRv...92..279B}. (In our simple estimate of $Q^*$ we do not have a size dependence, though.) Still, it is hard to see that $\gtrsim 10\ \mathrm{m\ s^{-1}}$ velocity collisions between equally-sized particles, even under these most favorable conditions, will not result in disruption; but this should of course really be tested by experiments.

On the other hand, it also seems sensible to pursue incremental growth scenarios which take place in quiescent (or low-$\alpha$) nebulae. Due to the relatively low effective $\alpha$-values for shear turbulence derived above, it is the radial drift motions that will provide the limits to growth. However, even a modest reduction of radial drift motion by a few factors due to collective effects may already be sufficient to prevent catastrophic collisions as particles reach $\mathrm{St} = 1$ \citep[see,][]{1993Icar..106..102C,2006Icar..181..572W}. Recall that fragmentation is easiest for nearly equal-size particles, which collide at very low velocities due to their systematic, nonturbulent motions. Moreover, the compound size distribution, which is a function of height, also determines whether collisions are beneficial to the growth; for example, if the size distribution is nearly monodisperse this will certainly favor growth in the nonturbulent cases. Since all these effects will vary with height, however, it is difficult to predict how these effects will unfold, and which parameters are key. Clearly, additional modeling is needed, where we may even combine these two different modes of turbulence since it is quite natural to expect that different physical processes operate at different heights \citep{2007ApJ...654L.159C}. However, incremental growth in the dense, particle-dominated midplanes of nonturbulent models then proceeds extremely rapidly \citep{1993Icar..106..102C,2000SSRv...92..295W} which is contrary to the evidence from meteorites and asteroids (see \citealt{2005ASPC..341..732C} or \citealt{2006mess.book..353C} for a discussion).

\subsection{Dust rim and matrix}
\begin{figure}
  \plotone{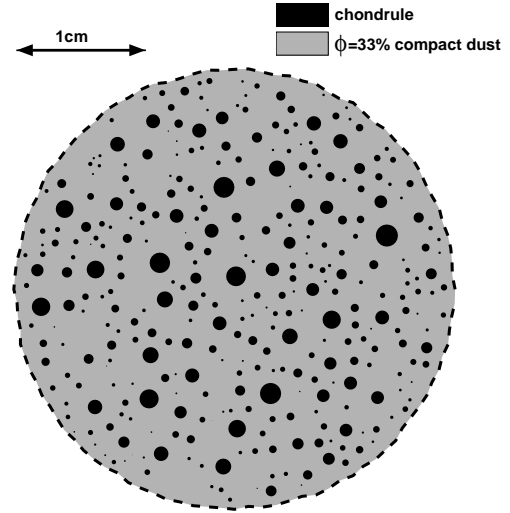}
  \caption{\label{fig:meteorite} A sketch of a cross-cut through a compound at the final state of our model. The compound contains two phases, present in equal proportion by mass: chondrules (black) and $\phi=0.33$ compact dust (grey). The cross-cut introduces a selection effect and shifts the chondrule size distribution to bigger chondrules. The chondrules are placed at random but a certain distance between the chondrules is preserved (see text).}
\end{figure}
\Fg{meteorite} provides an illustration of the internal structure of the objects obtained at the end of our simulation. In \fg{meteorite} it is assumed that each chondrule (black circles) is surrounded by a dust rim at least $\sim 40\%$ of the chondrule's mass, corresponding to the amount of dust accreted by individual chondrules (see \fg{hist}). This translates into an outer rim radius that is a factor of $1.3$ larger than that of the chondrule and is reflected in the inter-chondrule spacing of \fg{meteorite}. For the remainder the chondrules are positioned at random. (Note that \fg{meteorite} does not follow directly from the Monte Carlo collision model since we do not keep track of the positions of chondrules within compounds and cannot `reconstruct' a compound.) Furthermore, we have assumed the initial chondrule size distribution holds, but accounted for the bias a cross cut introduces to the observed structure \citep{1996M&PS...31..243E}. The chondrule size distribution is therefore skewed towards larger chondrules. 

However, the picture of a uniform 67\% porosity dust phase between chondrules contrasts with the meteoritic record. Here, the dust is compacted to a much larger extent and -- at least in the pristine CM chondrites but in others as well -- can be divided into two distinct components: the fine-grained, low porosity (10-20\%; \citealt{2006GeCoA..70.1271T}) rim that surrounds chondrules and the interstitial matrix material. \Fg{meteorite}, however, does not show this fine structure as the physical processes responsible for it were not modeled. Yet, the observed fact that chondrules in chondrites are separated by dust and not clustered together supports the main idea of this paper: that dust is accreted by chondrule-sized and perhaps somewhat larger particles (compounds) but not, \eg\ by planetesimals. How then did the rim-matrix distinction originate? Two scenarios can be envisioned: \textit{i)} a period of nebula dust sweep-up and compaction; or \textit{ii)} shock waves in the parent body.

The first scenario concerns a moderately intense turbulent environment (\ie\ high $\alpha$) in which chondrules are largely unable to stick, so that most of the dust is accreted by individual chondrules. These bouncing chondrules quickly compact each other's rims, while grazing collisions may also result in dust being partially stripped away or eroded from the rims. Presumably, a steady-state between rim accretion and erosion is established, where some of the dust is firmly attached to each chondrule and compacted, while another, more fluffy, component is continuously eroded off and reaccreted to the chondrule surfaces. Any of this latter, loosely bound phase which remains attached to chondrules at the point they are accreted to their parent planetesimal would be easily stripped away in the abrasive environment of the accreting planetesimal to become `matrix.' 
 
Alternatively, a much more gentle collisional environment may be considered in which big compounds form very quickly, and then continue to grow to planetesimal sizes. The dust is then primarily accreted by compounds, though, as we have argued in \se{canrun}, a significant proportion of the dust is always accreted by individual chondrules. As collisional energies stay low no identifiable rims are formed; and any rim signature might anyway easily abrade off on the parent planetesimal. In this scenario, the fine-grained dust rims might result from later processes on the parent body. Specifically, it has been suggested that shock waves through these planetesimals (caused, for example, by violent collisions with other planetesimals) will compact the dust \citep{2006GeCoA..70.1271T}. In the \citet{2006GeCoA..70.1271T} model the highest compaction of the dust takes place near the solid chondrule surface. Thus, it is only during the planetesimal stage that rims become distinguishable from the matrix. 

There are a number of constraints the rim formation mechanisms must satisfy. For instance, collisions must be energetic enough to compact the rims significantly to explain the high filling factors observed in chondrites. In the nebula formation scenario, therefore, more energetic collisions are required than provided by the model we present here. From the arguments given at the end of \se{collcomp}, we can estimate the velocities required to compact the accretion rims into nearly random closely packed configurations to be $10^{5/2}$ higher than the sticking velocity (\eqp{stick}), or $v \sim0.1\ \mathrm{km\ s^{-1}}$. Clearly, other compaction processes are needed than can be provided by collisions in turbulence, even between compounds approaching $\mathrm{St}=1$ in size. One possibility, given the preference for melting of chondrules by Mach 7 shock waves \citep{2002M&PS...37..183D,2005ASPC..341..873H}, is that the plausibly more numerous, more prevalent, weaker shocks which are consequently experienced even more routinely by particles, would provide this range of collisional velocities for chondrule-size particles and their fractal aggregates \citep{2006M&PS...41.1347C}. Another constraint is the rim-matrix distinction. In the parent body shock scenario it must therefore be shown that this distinction unambiguously results from these shocks. \citet{2006GeCoA..70.1271T} provide a qualitative idea on how this mechanism operates, and it would be desirable for this hypothesis to be backed-up more quantitatively by, \eg\ sophisticated numerical simulation.

\begin{figure}
  \plotone{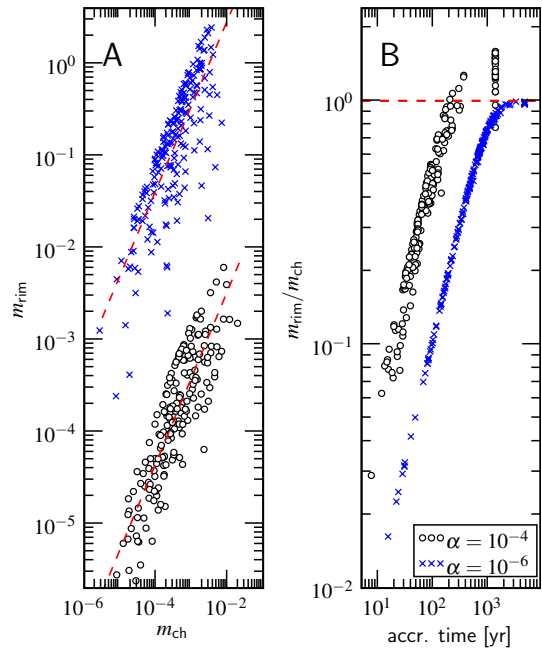}
  \caption{\label{fig:rimaccr} Model predictions for the thickness of the chondrule dust rim. The amount of dust accreted before the chondrule's incorporation into a compound determines the dust rim mass ($m_\mathrm{rim}$). (\textit{A}) Scatter plot of chondrule mass $m_\mathrm{ch}$ against $m_\mathrm{rim}$ for 200 chondrules, chosen randomly at the conclusion of the run. Two models are shown: $\alpha=10^{-4}$ and $\alpha=10^{-6}$ with the latter population being shifted by a factor of $10^3$ for clarity. The least-squares power-law fits are given by the dashed lines, which have exponents of 0.93 and 0.91, respectively. (\textit{B}) Dust-chondrule mass ratio ($y$-axis) at the time of its accretion into a compound ($x$-axis).}
\end{figure}
Yet a third observational constraint is the linear correlation between chondrule radius and rim thickness \citep{1992GeCoA..56.2873M,1997LPI....28.1071P}. 
\Fg{rimaccr} illustrates this point in the context of our accretion model. In \fg{rimaccr}A the chondrule-mass ($m_\mathrm{ch}$, $x$-axis) is plotted against the rim mass ($m_\mathrm{rim}$, $y$-axis) for the two models discussed in detail in \se{canrun}: crosses for the $\alpha=10^{-6}$ model and circles for the $\alpha=10^{-4}$ model. (In \fg{rimaccr}A the $\alpha=10^{-6}$ points are arbitrarily offset vertically by a factor of $10^3$ for reasons of clarity.) $m_\mathrm{rim}$ is defined as dust that is accreted by individual chondrules, before they become incorporated into a compound. The $m_\mathrm{ch}$-$m_\mathrm{rim}$ relation is shown for 200 chondrules, randomly selected from the initial distribution. The dashed lines show the best fit having slopes of $0.93$ and $0.91$, respectively. The near-linear trend of $m_\mathrm{rim}$ with chondrule mass is obvious but the spread is large, as seen in actual chondrites \citep{1992GeCoA..56.2873M}. \Fg{rimaccr}B shows the accretion history of these compounds: the mass ratio, $m_\mathrm{rim}/m_\mathrm{ch}$, is plotted ($y$-axis) against the time at which the chondrule is swept up by a compound. The `$\alpha=10^{-4}$ chondrules' lie to the left of the `$\alpha=10^{-6}$ chondrules,' reflecting their shorter collision times. The initially linear trend breaks down at later times as the density of free-floating dust decreases. Although many processes contribute to the spread in the data points of \fg{stats}A -- for example, differences in velocity field (linear/square-root regime) during the simulation and the bouncing history of chondrules -- the stochasticity in the chondrule-compound accretion time is the main contributor. Note also the pile-up of particles in the $\alpha = 10^{-4}$ model near $t \sim 10^3\ \mathrm{yr}$, the final time of the simulation: these are the chondrules that remained single during the entire simulation.

The relation of rim thickness with chondrule size can be naturally understood as the outcome of a nebula accretion process (\citealt{1998Icar..134..180M,2004Icar..168..484C}; \fg{rimaccr}). The observed linear relationship in chondrites therefore suggests this relationship should somehow have survived further processing. As dust rim accretion in the violent collisional environment differs from the non-fragmentation environment in which our simulations are performed, it still remains to be shown that the linear relationship is maintained after fragmentation/erosion sets in. Alternatively, if the imprints of nebula dust-accretion are destroyed during parent body accretion, a different mechanism must explain the observed relationship. 

Future work -- \eg\ experimental work on rim-chondrule size ratios and more advanced theoretical models -- must determine which of the two scenarios described above is more likely. Dust fragmentation and additional compaction mechanisms may be included into the present model. Increasingly energetic collisions (when compounds grow towards the $\mathrm{St}=1$ barrier) may disrupt compound objects without stripping the rims entirely off of individual chondrules, and in doing so may compact the surviving fine grained rims further than the $\phi_\mathrm{cd}=0.33$ limit we have adopted in this study. 
Also, size distributions in the fine-grained component might also allow a greater degree of packing than in our models and \citet{2004PhRvL..93k5503B} expect, in which the grains are all monodisperse.

\section{Summary}
\label{sec:summ}
We have investigated a chondrule-dust aggregation mechanism in which the fine-grained dust acts as the glue that allows chondrules to stick. We argue that the energy in collisions is sufficient to compress directly accreted material, which initially has a porous `fairy-castle' structure, into a more compact state having a porosity that is roughly 67\% (based on compaction measurements by \citealt{2004PhRvL..93k5503B} and theoretical arguments). We have applied this model to a variety of questions regarding the meteoritic record: the relation of individual chondrules to their fine-grained dust rims, the internal structure of the chondrites, and the ability of growth by sticking to surpass the meter-size barrier. This study only starts to address these questions; more sophisticated models are needed to answer detailed questions on the structure of the meteorites.

We find that porous accretion rims do indeed cushion collisions and facilitates growth to compound objects containing many rimmed chondrules, but this growth is limited to $30-100\ \mathrm{cm}$ radius objects under the most favorable conditions. This is because the chondrule component sweeps up all the local dust in a short time ($10^2 - 10^4\ \mathrm{yr}$, depending on nebular location) and these compounds experience higher relative velocities during their growth stage. Subsequent collisions merely pack the existing rims down further, so that the system ultimately reaches a dead-end steady state where collisions only result in bouncing, or possibly disruption. Other conclusions from this study are:

\begin{itemize}
\item Compound growth works best in a quiescent environment (high gas density, low $\alpha$ values). In a more violent collisional environment ($\alpha \sim 10^{-3..-2}$) it is difficult to accrete dust fractally on chondrules surfaces and the energetic collisions between compounds quickly compact the remainder such that collisional growth is quickly terminated. 
\item The importance of the other parameters on the accretion process is mostly minor. The radial location does not affect the final growth of the compounds, although timescales are longer at larger $R$. Ice, rather than silicate, as the sticking agent will lead to bigger compounds (but we note icy grains do not dominate the meteoritic record).
\item In no single model do compounds grow to planetesimal sizes. Either turbulent or systematic velocities are too high for the porous dissipation mechanism, the $\mathrm{St}=1$ size being the bottleneck; but we may have over-estimated the systematic velocities in this study by not accounting for particle collective effects in low turbulence nebulae.
\item We anticipate that the dust accreted by individual chondrules -- before chondrules coagulate into compounds -- will finally end up as the chondrule rim. In strong turbulent models this fraction is very high, but it remains significant (tens of percents) even in the collisionally gentle models. 
\item However, at the current state of the art of this model, fine-grained accretion rims have a porosity significantly larger than seen in actual rims. Other compaction processes are not hard to envision, such as higher velocity collisions by larger mass compound objects, or nebula shock waves (peripheral to those energetic enough to melt chondrules). These remain to be modeled.
\item When we define the rim as the dust swept up by individual chondrules, we find very good agreement with the nearly linear (average) correlation between rim thickness and underlying chondrule radius seen in CM and CV chondrites \citep{1992GeCoA..56.2873M,1997LPI....28.1071P}.
\end{itemize}

Future work will focus on two aspects of our coagulation model:
\begin{itemize}
  \item An improvement of the collisional physics, \ie\ including fragmentation as a collisional outcome for velocities above $\sim \mathrm{m\ s^{-1}}$; and a refinement of the characterization of the compound structure, \eg\ to allow dust to compact to higher filling factors.
\item Inclusion of a proper description of the vertical structure of the nebula, \ie\ taking account of phenomena such as settling, collective effects, and shear turbulence. Especially, the transition from global- ($\alpha$) to shear-turbulence is important, and in future work we will address this critical junction.
\end{itemize}

\acknowledgements 
We thank J\"urgen Blum, Carsten Dominik, Alan Rubin, Marco Spaans and John Wasson for helpful conversations. C.W.O. acknowledges a grant from the Netherlands Organisation for Scientific Research (NWO). J.N.C.'s contributions were supported by a grant from NASA's Origins of Solar Systems program. We thank the anonymous referee for comments that helped clarifying the paper.
\bibliographystyle{apj}
\bibliography{ms}
\end{document}